\documentclass[aps,superscriptaddress,twocolumn]{revtex4-2}
\usepackage[utf8]{inputenc}
\usepackage{amsthm,amssymb,amsmath}
\usepackage{graphicx,xcolor}
\usepackage[colorlinks=true,linkcolor=blue,citecolor=blue,urlcolor=blue,linktocpage=true]{hyperref}

\usepackage{microtype}
\usepackage{booktabs}
\usepackage{float}
\usepackage{multirow}
\usepackage{siunitx}
\usepackage[vlined,ruled]{algorithm2e}
\usepackage{tabulary}

\begin{document}
\title{Improving precision scaling via backaction-evading continuous measurement in a driven-dissipative Kerr parametric oscillator}

\author{Cheng Zhang}
\affiliation{Research Center for Quantum Physics and Technologies, Inner Mongolia University, Hohhot 010021, China}
\affiliation{School of Physical Science and Technology, Inner Mongolia University, Hohhot 010021, China}

\author{Xinhui Cui}
\affiliation{Research Center for Quantum Physics and Technologies, Inner Mongolia University, Hohhot 010021, China}
\affiliation{School of Physical Science and Technology, Inner Mongolia University, Hohhot 010021, China}

\author{Jiaying Pan}
\affiliation{Research Center for Quantum Physics and Technologies, Inner Mongolia University, Hohhot 010021, China}
\affiliation{School of Physical Science and Technology, Inner Mongolia University, Hohhot 010021, China}

\author{Xin-Qi Li}
\affiliation{Research Center for Quantum Physics and Technologies, Inner Mongolia University, Hohhot 010021, China}
\affiliation{School of Physical Science and Technology, Inner Mongolia University, Hohhot 010021, China}

\author{Mauro Cirio}
\affiliation{Graduate School of China Academy of Engineering Physics, Haidian District, Beijing, 100193, China}

\author{Pengfei Liang}
\email{pfliang@imu.edu.cn}
\affiliation{Research Center for Quantum Physics and Technologies, Inner Mongolia University, Hohhot 010021, China}
\affiliation{School of Physical Science and Technology, Inner Mongolia University, Hohhot 010021, China}

\date{\today}
\begin{abstract}
Dissipative phase transitions in the driven-dissipative Kerr parametric oscillator offer a promising route for realizing criticality-enhanced quantum sensing based on continuous measurements. However, achieving such enhancement through realistic measurement schemes remains an outstanding challenge. 
Here, we extend the backaction-evasion strategy introduced in our earlier work for the Gaussian linear case [arXiv:2511.22248 (2025)] to analyze how the quantum and classical Fisher information scale with the Kerr nonlinearity at dissipative critical points. Our results show that backaction-evading homodyne monitoring achieves enhanced photon-number scaling that surpasses the standard quantum limit, and significantly outperforms alternative protocols such as continuous photon counting. As an additional methodological contribution, we also implement and benchmark time-discrete approximation schemes with improved statistical convergence properties. We use these methods to compute the classical Fisher information for continuous homodyne detection, and demonstrate that they provide efficient access to this quantity near dissipative critical points, thereby extending the reach of existing methods.
\end{abstract}

\maketitle

\section{Introduction}

Continuous measurements provide a versatile tool for extracting useful information from a quantum system while weakly disturbing it in a controlled, non-projective manner~\cite{Petruccione,Gardiner,Wiseman_Milburn_2009}. In quantum information processing, continuous measurements are central to quantum state preparation and stabilization~\cite{Sayrin2011, Vijay2012, Riste2013}, quantum error correction~\cite{Ahn2000, Kapit2016}, and measurement-based quantum control~\cite{Blais2004, Minev2019}. In quantum metrology, continuous monitoring of the emission field of quantum-optical sensors enables real-time parameter estimation~\cite{PhysRevA.102.063716,amorosbinefa2025trackingtimevaryingsignalsquantumenhanced}, quantum filtering~\cite{Wiseman1994, Bouten2007}, and feedback controls~\cite{Wiseman1994, Doherty2000, Zhang2017}. Compared with strong projective measurements, continuous measurements track 
the conditional quantum state along individual measurement records, allowing one to tune the balance between collected information and the resulting back-action on the system \cite{Fuchs1996, Jacobs2006}. 
These features are particularly attractive for quantum sensing schemes in which the system of interest is continuously coupled to a bosonic environment, such as in cavity- and circuit-QED platforms~\cite{Blais2004, Blais2021}. 

Quantum critical systems have recently emerged as promising sensors because their enhanced susceptibility near a phase transition can amplify the response to small external perturbations~\cite{Zanardi2008, Invernizzi2008, Rams2018}. As a consequence, driven-dissipative systems supporting dissipative phase transitions constitute a natural avenue to combine the advantages of quantum criticality and continuous monitoring. For example, in quantum-optical settings, 
the emitted light carries information about the system while the associated measurement backaction can strongly affect the achievable sensitivity~\cite{Albarelli_2017,Albarelli2018restoringheisenberg,PhysRevLett.125.200505,k7nk-lrwd}. However, how to extract this information to approach the ultimate precision limit set by the quantum Cram\'er-Rao bound remains an open question, particularly for continuous measurements limited by 
current experimental capabilities. 

A general step forward in analyzing this issue was provided in Ref.~\cite{PhysRevX.13.031012}, which proposed a sensor-decoder cascaded scheme capable of nearly saturating the quantum Cram\'er-Rao bound. Nevertheless, when applied to critical quantum sensing, implementing the decoder may require unconventional reservoir engineering, which severely undermines the practicality of this scheme in such a scenario. An alternative approach, applicable to critical sensing with quantum-optical sensors featuring dissipative critical points, was reported in our earlier work~\cite{zhang2025enhancinginformationretrievalquantumoptical}. Our approach exploits an interesting interplay between measurement backaction and quantum criticality, and enables optimal temporal scaling of the attainable precision in certain idealized regimes.

In this work, we extend the analysis in Ref.~\cite{zhang2025enhancinginformationretrievalquantumoptical} of the open Kerr parametric oscillator (KPO), a quantum-optical driven-dissipative system exhibiting tunable dissipative phase transitions~\cite{PhysRevLett.133.040801} accompanied by $\mathbb{Z}_2$ symmetry breaking. The KPO serves as a minimal, and yet experimentally relevant platform for quantum critical sensing, suitable for both strong projective measurements~\cite{DiCandia2023,PhysRevLett.133.040801} and continuous monitoring schemes~\cite{zhang2025enhancinginformationretrievalquantumoptical}. In Ref.~\cite{zhang2025enhancinginformationretrievalquantumoptical}, we analyzed this setup subject to continuous measurements in the idealized limit of vanishing Kerr nonlinearity. In this regime, it is possible to impose a phase-matching condition which makes continuous homodyne detection optimal in terms of the temporal scaling of the attainable precision, thereby outperforming other measurement schemes.

However, it is not clear whether, under the same phase-matching condition, continuous homodyne detection still retains its advantage for a finite Kerr nonlinearity. This question is particularly relevant because a nonzero Kerr nonlinearity is required to regularize the dynamics in the $\mathbb{Z}_2$ symmetry-broken phase, which makes it an essential metrological resource in this setting.  
As a consequence, estimating the scaling of the achievable precision with respect to the Kerr nonlinearity is crucial to assess whether continuous homodyne detection can offer 
a genuine metrological advantage over alternative schemes such as continuous photon counting. To enhance the significance of this approach, it is further relevant to analyze whether such a scaling behavior is optimal at dissipative critical points.

This work provides a twofold contribution to these open questions. First, building on the general formalism in Ref.~\cite{Platen}, we implement and benchmark several numerical methods for computing the classical Fisher information of continuous homodyne detection in the vicinity of dissipative critical points. This technical contribution extends the scope of the existing literature~\cite{PhysRevA.91.012118,Albarelli2018restoringheisenberg} and may find applications in other theoretical analyses of quantum sensing based on continuous homodyne detection. Second, we apply these methods to the KPO and extract the critical exponents governing the scaling behaviors of both the quantum and classical Fisher information through a finite-size scaling analysis. Our results show that the homodyne backaction-evading configuration does offer enhanced precision scaling that exceeds the standard quantum limit, although it does not saturate the bound set by the quantum Fisher information. Theoretical values for the relevant critical exponents are summarized in Table~\ref{tab:exponents}, with their meanings given in Sec.~\ref{sec:FSStheory}. 

The rest of the paper is organized as follows. In Sec.~\ref{sec:sensordyn}, we introduce the general framework of quantum sensing based on continuous measurements, and in Sec.~\ref{sec:QRCB}, the corresponding quantum Cram\'er-Rao bound. In Sec.~\ref{sec:CFInum} we present the numerical methods for computing the classical Fisher information for both continuous homodyne detection and photon counting. Importantly, we implement several time-discrete approximations with improved statistical convergence properties that extend the existing literature~\cite{PhysRevA.91.012118,Albarelli2018restoringheisenberg}, and we demonstrate their efficiency near dissipative critical points. In Sec.~\ref{sec:KPO} we introduce the KPO model and summarize some important results from our earlier work~\cite{zhang2025enhancinginformationretrievalquantumoptical} that are relevant to the scaling analysis detailed in Sec.~\ref{sec:scalinganalysis}. We conclude and discuss  in Sec.~\ref{sec:conclusions}.

\section{Quantum sensing based on continuous measurements}\label{sec:QSCM}
We begin by outlining the general framework of quantum sensing based on continuous measurements in Secs.~\ref{sec:sensordyn} and~\ref{sec:QRCB}. Throughout, we focus on estimating a single parameter, though the numerical methods presented in Sec.~\ref{sec:CFInum} can be extended to multi-parameter estimation without significant difficulty. 

\subsection{Sensor dynamics under continuous monitoring}\label{sec:sensordyn}
We assume that the parameter of interest, $\theta$, is encoded in the Hamiltonian $H_\theta$ of a quantum sensor that interacts with a Markovian quantum environment consisting of a continuum of bosonic modes. The effects of the environment can be modeled as quantum white noise in terms of the bosonic noise operator $b(t)$~\cite{Petruccione,Gardiner,Wiseman_Milburn_2009}, which satisfies the commutation relation $[b(t), b^\dagger(t')] = \delta(t-t')$. The interaction between the sensor and the environment at time $t$ is described by the unitary $U_t = \exp[-iH_\theta dt + i(c b^\dagger(t) - b(t)c^\dagger)dt]$, where the jump operator $c$ acts on the Hilbert space of the sensor.  At time $t_n = n\,dt$ ($n\in\mathbb{N}$), the state of the joint system (the sensor and the environment) can be expressed as 
\begin{equation}
|\Psi_\theta\rangle = U_{t_{n-1}}\cdots U_{t_0}|\psi_\text{S}(0)\rangle\otimes|0_{n-1},\cdots,0_0\rangle, 
\end{equation}
where $|0_i\rangle$ ($i=0,\cdots,n-1$) denotes the vacuum of the noise operator $b(t_i)$ and $|\psi_\text{S}(0)\rangle$ is the initial state of the sensor. The reduced state of the sensor, defined by 
tracing out all environmental degrees of freedom and denoted as $\varrho = \operatorname{Tr}_\text{E}[|\Psi_\theta\rangle\langle\Psi_\theta|]$, evolves according to the Lindblad master equation
\begin{equation}\label{eq:me}
d\varrho/dt = \mathcal{L}\varrho \equiv -i[H_\theta, \varrho] + \mathcal{D}[c]\varrho,
\end{equation}
where the dissipator $\mathcal{D}[c]\varrho = c\varrho c^\dagger - \frac12 \{c^\dagger c,\varrho\}$ accounts for dissipation into the Markovian environment. 

In this setting, a time-local continuous measurement corresponds to sequential projective measurements on the environment, formally described by a collection of projection operators $\{P_n:n=0,1,\cdots\}$ where $P_n$ acts on the Hilbert space of the noise mode $b(t_n)$. We focus on two widely studied continuous measurement schemes in quantum optics. 

The first is continuous homodyne detection, in which at time $t_n$ the field quadrature $X_{n,\varphi} \equiv (B_ne^{-i\varphi} + B_n^\dagger e^{i\varphi})/\sqrt{2}$ is measured, where $B_n \equiv b(t_n)\sqrt{dt}$ satisfies the canonical commutation relation $[B_n, B_n^\dagger] = 1$. In this case, the projection operator is $P_n = |x_{n,\varphi}\rangle\langle x_{n,\varphi}|$, where $|x_{n,\varphi}\rangle$ denotes the eigenstate of $X_{n,\varphi}$ with eigenvalue $x_{n,\varphi}$. 
The homodyne phase $\varphi$ can thus be tuned to select the measured quadrature. The sensor state $\rho$ becomes conditioned on the past measurement outcomes and obeys the stochastic master equation (SME)~\cite{Wiseman_Milburn_2009} 
\begin{equation}\label{eq:sme_hd}
d\rho = \mathcal{L}\rho\,dt + \sqrt{\eta}\mathcal{H}[c_\varphi]\rho\,dw,  
\end{equation}
where the superoperator $\mathcal{H}[c_\varphi]\rho = c_\varphi\rho + \rho c^\dagger_\varphi - \langle c_\varphi+c^\dagger_\varphi\rangle_\rho \rho$, with $\langle \cdot\rangle_\rho = \text{Tr}[\cdot\rho]$ and $c_\varphi = ce^{-i\varphi}$, captures the quantum measurement backaction, $dw$ is a Wiener increment satisfying $\mathbb{E}[dw]=0$ and $\mathbb{E}[dw^2]=dt$, and the detection efficiency $\eta\in[0,1]$  accounts for experimental imperfections~\cite{Wiseman_Milburn_2009}, such as dark noise in photodetectors and unmonitored photon loss channels.  The measurement outcome at time $t$ is given by the photocurrent  
\begin{equation}\label{eq:photocurrent}
dy_t = \sqrt{\eta}\langle c_\varphi+c^\dagger_\varphi\rangle_\rho\,dt + dw. 
\end{equation}
Here, the first term is the deterministic contribution arising from the 
projection of the environmental state on $|x_{n,\varphi}\rangle$, and effectively leading to continuous monitoring of the sensor quadrature $c_\varphi+c^\dagger_\varphi$.

The other measurement scheme that we analyze is continuous photon counting. In this scenario, the projection at time $t_n$ is either $P_n = |0_n\rangle\langle0_n|$ or $P_n = |1_n\rangle\langle1_n|$, where $|0_n\rangle$ and $|1_n\rangle$ are the Fock states of the mode $b(t_n)$. As a result, the measurement outcome within the time interval $(t,t+dt)$ is given by a Poisson increment $dN_t$ taking the value $0$ or $1$.  Specifically, $dN_t=1$ ($dN_t=0$) corresponds to a photon click (no-click) event, and the probability of a click occurring is $\mathbb{E}[dN_t] = \eta\langle c^\dagger c\rangle_\rho dt$.  The SME describing the sensor dynamics is
\begin{equation}\label{eq:sme_pc}
d\rho = \mathcal{L}\rho\,dt + \eta\bigl(\langle c^\dagger c\rangle_\rho\rho - c\rho c^\dagger\bigr)dt + \left( \frac{c\rho c^\dagger}{\langle c^\dagger c\rangle_\rho} - \rho \right)dN_t, 
\end{equation}
where the last term, which is independent of the detection efficiency, introduces discontinuous jumps in the dynamics, corresponding to the measurement backaction when a photon is recorded by the detector. When no photon is detected, $dN_t=0$, the dynamics is still affected by measurement backaction due to the presence of the second term~\cite{Wiseman_Milburn_2009}. 

In the next subsection, we introduce the quantum Cram\'er-Rao bound, which sets the ultimate precision limit for parameter estimation in sensing protocols based on continuous measurements.

\subsection{Quantum Cram\'er-Rao bound for sensing schemes based on continuous measurements}\label{sec:QRCB}
We denote by $\mathbf{Y}_t$ the collection of measurement outcomes up to time $t$. As shown in the previous section, for homodyne detection it is defined as $\mathbf{Y}_t = \{dy_s: 0\le s \le t\}$, while for photon counting it is $\mathbf{Y}_t = \{dN_s: 0\le s \le t\}$. An estimator for the parameter $\theta$ is formally defined as a random variable $\hat\theta(\mathbf{Y}_t^{(1)},\cdots,\mathbf{Y}_t^{(K)})$, where $K$ is the number of measurement repetitions. For sufficiently many repetitions, $K\gg1$, the quantum Cram\'er-Rao bound sets a lower bound on the sensitivity for estimating $\theta$,  quantified by the variance $\operatorname{Var}[\hat\theta]$, through the inequality
\begin{equation}\label{eq:QRCB}
\operatorname{Var}[\hat\theta] \ge \frac{1}{K I} \ge \frac{1}{K I_\text{G}}. 
\end{equation}
Here, $I$ denotes the classical Fisher information for a generic measurement performed on the joint state $|\Psi_\theta\rangle$. In what follows, whenever possible, we append a subscript to $I$ to distinguish the classical Fisher information for different continuous measurement schemes, 
e.g., $I_\text{hd}$ for homodyne detection and $I_\text{pc}$ for photon counting. 

The quantity $I_\text{G}$ refers to the global quantum Fisher information defined in terms of $|\Psi_\theta\rangle$ as~\cite{PhysRevLett.112.170401,PRXQuantum.3.010354,PhysRevX.13.031012}
\begin{equation}\label{eq:IGdef}
I_\text{G} = -4\partial_{\delta}^2 \mathcal{F}_\text{G}(\theta,\theta+\delta) \big\vert_{\delta=0}, 
\end{equation}
where $\mathcal{F}_\text{G}(\theta_1,\theta_2) = \lvert\langle\Psi_{\theta_1}|\Psi_{\theta_2}\rangle\rvert$ is the quantum fidelity of the joint state. Defining the operator $\mu_{\theta_1,\theta_2} = \operatorname{Tr}_\text{E}\bigl[|\Psi_{\theta_1}\rangle\langle\Psi_{\theta_2}|\bigr]$, it can be shown that $\mathcal{F}_\text{G}(\theta_1,\theta_2) = \lvert\operatorname{Tr}_\text{S}[\mu_{\theta_1,\theta_2}]\rvert$, and that $\mu_{\theta_1,\theta_2}$ satisfies the time-local generalized master equation~\cite{PhysRevLett.112.170401,PRXQuantum.3.010354,PhysRevX.13.031012}
\begin{equation}\label{eq:muEq}
\frac{d\mu_{\theta_1,\theta_2}}{dt} = -i(H_{\theta_1}\mu_{\theta_1,\theta_2} - \mu_{\theta_1,\theta_2} H_{\theta_2}) + \mathcal{D}[c]\mu_{\theta_1,\theta_2}. 
\end{equation}
This master equation can then be numerically integrated to obtain $\mu_{\theta_1,\theta_2}$, from which $I_\text{G}$ can be calculated by approximating the derivatives in Eq.~(\ref{eq:IGdef}) in terms of finite differences.

It is worth noting that a tighter bound can be derived if one considers measurement schemes performed solely on the environment, such as the continuous homodyne detection and photon counting outlined in the previous subsection. Restricting to those measurements, the quantum Cram\'er-Rao bound~(\ref{eq:QRCB}) can be tightened by using the quantum Fisher information of the reduced state of the environment $\rho_\text{E}(\theta) = \operatorname{Tr}_\text{S}\bigl[|\Psi_{\theta}\rangle\langle\Psi_{\theta}|\bigr]$, defined as~\cite{ltfw-4fwn}
\begin{equation}\label{eq:IE}
I_\text{E} = -4\partial_{\delta}^2 \mathcal{F}_\text{E}(\theta,\theta+\delta) \big\vert_{\delta=0}, 
\end{equation}
where $\mathcal{F}_\text{E}(\theta_1,\theta_2)$ is the quantum fidelity of $\rho_\text{E}(\theta)$, defined as
\begin{equation}
\mathcal{F}_\text{E}(\theta_1,\theta_2) = \operatorname{Tr}_\text{S}\left( \sqrt{ \sqrt{\rho_\text{E}(\theta_1)} \rho_\text{E}(\theta_2) \sqrt{\rho_\text{E}(\theta_1)} } \right). 
\end{equation}
By definition, $I_\text{E}\le I_\text{G}$. A useful identity,  
proven in Ref.~\cite{ltfw-4fwn}, is $\mathcal{F}_\text{E}(\theta_1,\theta_2) = \operatorname{Tr}_\text{S}[\sqrt{\mu_{\theta_1,\theta_2}\mu_{\theta_1,\theta_2}^\dagger}]$, which allows one to calculate $I_\text{E}$ by numerically solving Eq.~(\ref{eq:muEq}). In this work, we focus exclusively on the quantum Fisher information $I_\text{G}$, as it is more tractable numerically, especially in certain limiting regimes. However, 
the results for $I_\text{G}$ presented later in Sec.~\ref{sec:scalinganalysis} from finite-scaling analysis equally apply also to $I_\text{E}$, since the latter exhibits the same long-time behavior as $I_\text{G}$.

In parallel, the classical Fisher information is defined as
\begin{equation}\label{eq:Fdef}
I = \mathbb{E}\bigl[(\partial_\theta \ln p(\mathbf{Y}_t \vert \theta))^2\bigr], 
\end{equation} 
where the conditional distribution $p(\mathbf{Y}_t \vert \theta)$ denotes the probability of obtaining the record $\mathbf{Y}_t$ conditioned on $\theta$. Since $p(\mathbf{Y}_t \vert \theta)$ is a distribution in a high-dimensional space (whose dimensionality grows linearly with $t$), it is generally difficult to access this quantity analytically. This  
raises the need for numerically efficient methods for calculating the classical Fisher information for continuous measurements, particularly in certain difficult regimes in quantum critical sensing. This  constitutes the major challenge we address in the subsequent subsection.

\begin{figure*}[t!]
\includegraphics[clip,width=18cm]{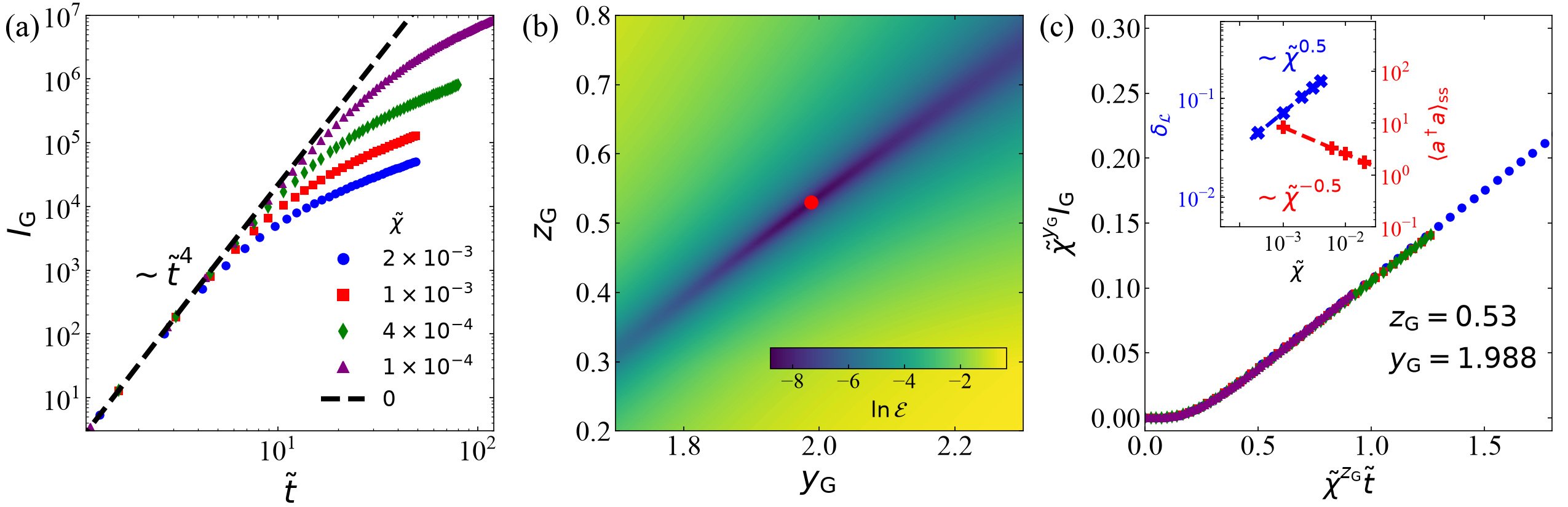}
\caption{(a) The global quantum Fisher information $I_\text{G}$ as a function 
of time at the dissipative critical point $\tilde{\epsilon}=\tilde{\epsilon}_c(\tilde{\omega})$, for $\tilde{\omega}=1.0$.
The dashed line corresponds to the Gaussian case $\tilde{\chi}=0$, while colored markers correspond to different non-zero values of the Kerr nonlinearity $\tilde{\chi}\neq0$. (b) Minimization of the scaled collapse error $\mathcal{E}$ to determine the optimal values of the exponents $y_\text{G}$ and $z_\text{G}$ (marked by the red dot). (c) Optimal collapse of the data from panel (a) using the scaling ansatz for $I_\text{G}$ in Eq.~(\ref{eq:Iansatz}), with the exponents set to their optimal values (black texts) extracted in panel (b). The inset shows the Liouvillian gap $\delta_\mathcal{L}$ (crosses) and the steady-state photon number 
$\langle a^\dagger a\rangle_\text{ss}$ (plus symbols) as functions of $\tilde{\chi}$ at this critical point, along with the corresponding power-law fits (dashed lines). From fitting, the dynamical exponent $z=0.5$ (blue text) and the scaling dimension $\Delta_{a^\dagger a} = -0.5$ (red text) are extracted. 
}\label{fig:QFIfig}
\end{figure*}

\subsection{Numerical methods for calculating the classical Fisher information}\label{sec:CFInum}

The logarithmic derivative in Eq.~(\ref{eq:Fdef}) can be treated
using finite differences. This strategy has been used for the calculation of the classical Fisher information for continuous photon counting~\cite{PRXQuantum.3.010354,Ilias2024}. However, under homodyne monitoring, the sensor dynamics becomes diffusive, limiting the reliability of this option.  

In contrast, in Ref.~\cite{PhysRevA.87.032115}, Gammelmark and M\o lmer showed that the classical Fisher information for both continuous homodyne detection and continuous photon counting can be represented as
\begin{equation}\label{eq:Ftau}
I = \mathbb{E}\bigl[\operatorname{Tr}[\tau]^2\bigr],  
\end{equation}
where $\tau$ is an auxiliary matrix of the same dimension as the density operator $\rho$. Comparing Eq.~(\ref{eq:Ftau}) with the definition in Eq.~(\ref{eq:Fdef}) one finds that $\operatorname{Tr}[\tau] = \partial_\theta \ln p(\mathbf{Y}_t \vert \theta)$. As a consequence, while equivalent to the original definition, the representation in Eq.~(\ref{eq:Ftau}) avoids direct calculation of logarithmic derivatives. The dynamics of $\tau$ is determined by an It\^o stochastic differential equation, whose specific form is given in the next section as Eq.~(\ref{eq:tau_hdeq}) for homodyne detection and as Eq.~(\ref{eq:tau_pceq}) for photon counting.

\subsubsection{Continuous homodyne detection}\label{sec:hd}
For homodyne detection, the auxiliary matrix $\tau$ evolves according to the It\^o differential equation~\cite{Albarelli2018restoringheisenberg}
\begin{equation}\label{eq:tau_hdeq}
d\tau = \mathcal{L}\tau\,dt + \mathcal{M}\rho\,dt + \sqrt{\eta}\mathcal{H}[c_\varphi]\tau\,dw,  
\end{equation}
where we have introduced the superoperator $\mathcal{M}\rho = -i[\partial_\theta H_\theta, \rho]$. Because the conditional state $\rho$ appears explicitly on the right-hand side of this equation, Eqs.~(\ref{eq:sme_hd}) and (\ref{eq:tau_hdeq}) must be solved simultaneously to determine $\tau$. 

We now rewrite Eqs.~(\ref{eq:sme_hd}) and (\ref{eq:tau_hdeq}) in a more compact vector form. To this end, we introduce the vector $\Theta = (\rho, \tau)^\intercal$, and combine these two equations into a single vector equation
\begin{equation}\label{eq:eq_Theta}
d\Theta = \boldsymbol{\mathcal{A}}(\Theta)\,dt + \boldsymbol{\mathcal{B}}(\Theta)\,dw, 
\end{equation}
where the superoperators $\boldsymbol{\mathcal{A}}$ and $\boldsymbol{\mathcal{B}}$ are defined as
\begin{equation}
\boldsymbol{\mathcal{A}}(\Theta) = \begin{pmatrix}
\mathcal{L}\rho \\
\mathcal{L}\tau + \mathcal{M}\rho 
\end{pmatrix},~\boldsymbol{\mathcal{B}}(\Theta) = \begin{pmatrix}
\sqrt{\eta}\mathcal{H}[c_\varphi]\rho \\
\sqrt{\eta}\mathcal{H}[c_\varphi]\tau
\end{pmatrix}, 
\end{equation}
which, respectively, can be interpreted as the drift and diffusion coefficients in the context of It\^o stochastic processes~\cite{Platen}. 

Numerical simulations of diffusive It\^o stochastic processes, such as the one defined in Eq.~(\ref{eq:eq_Theta}), require the use of their time-discrete approximations. In this context, two approximation schemes widely used in the literature are the Euler-Maruyama and the Milstein methods~\cite{PhysRevA.91.012118,Albarelli2018restoringheisenberg}. However, these methods have primarily been applied to sensor models that do not involve dissipative phase transitions (e.g., atomic magnetometry~\cite{Albarelli2018restoringheisenberg,Albarelli_2017}, and Rabi frequency estimation via a single atom~\cite{PhysRevA.94.032103}), raising concerns about their reliability and efficiency near dissipative critical points. 

In fact, as demonstrated in Appendix~\ref{sec:numapprox}, both methods fail to provide efficient access to the classical Fisher information for continuous homodyne detection in this challenging regime. This limitation motivates approximation schemes with stronger convergence properties. To address this challenge, we implement and benchmark two additional methods, based on the systematic framework developed in Ref.~\cite{Platen}. These developments extend beyond the scope of Refs.~\cite{PhysRevA.91.012118,Albarelli2018restoringheisenberg} and may prove useful in other sensing scenarios involving continuous homodyne detection where accurate evaluation of the classical Fisher information is essential. However, given the rather technical nature of this contribution, we provide the detailed analysis in Appendix~\ref{sec:numapprox} to maintain a concise and focused presentation.

The case $\eta=1$ deserves special discussion. In this case, the SMEs in Eqs.~(\ref{eq:sme_hd}) and (\ref{eq:tau_hdeq}) both reduce to stochastic Schr\"odinger equations (SSEs), since both the density operators $\rho$ and $\tau$ can be expressed in terms of pure states as~\cite{Albarelli2018restoringheisenberg}
\begin{equation}\label{eq:rhotau_pure}
\rho = |\psi\rangle\langle\psi|,~~~\tau = |\phi\rangle\langle\psi| + |\psi\rangle\langle\phi|. 
\end{equation}
We also provide the explicit constructions of the new approximation schemes in terms of these conditional pure states in Appendix~\ref{sec:sse_hd}. Importantly, the reduction from density operators to pure states brings significant numerical efficiency, because the complexity of matrix multiplications in the SMEs in Eqs.~(\ref{eq:sme_hd}) and (\ref{eq:tau_hdeq}) is $\mathcal{O}(N^3)$, while it is $\mathcal{O}(N^2)$ in the SSEs in Eqs.~(\ref{eq:sse_hd}) and (\ref{eq:phi}), where $N$ denotes the dimension of the matrices in these stochastic equations. In our numerical practice, we find that the SMEs are typically ten times slower than the SSEs per trajectory, making the calculation of the classical Fisher information for $\eta<1$ near dissipative critical points computationally unaffordable. For this reason, we restrict our discussion to the case of ideal detection in the scaling analysis which we present 
in Sec.~\ref{sec:scalinganalysis}, and leave the treatment of the non-ideal case for future work. A possible solution to 
this issue is given in Ref.~\cite{Ilias2024} and consists in 
unraveling the unmonitored channel to derive the corresponding SSE and  consequently averaging over those unmonitored events to obtain the conditional distribution $p(\mathbf{Y}_t \vert \theta)$. However, to the best of our understanding, it is not clear whether this approach can also be applied to the case of continuous homodyne detection.

\subsubsection{Continuous photon counting}\label{sec:pc}
For photon counting, the evolution of $\tau$ is governed by
\begin{equation}\label{eq:tau_pceq}
\begin{aligned}
d\tau &= \mathcal{L}\tau dt + \eta\bigl(\langle c^\dagger c\rangle_\rho\tau - c\tau c^\dagger\bigr)dt + \mathcal{M}\rho dt \\
&~~~ + \left(\frac{c\tau c^\dagger}{\langle c^\dagger c\rangle_\rho} - \tau\right) dN_t. 
\end{aligned}
\end{equation}
Owing to the discrete nature of the Poisson increment $dN_t$, individual trajectories of both $\rho$ and $\tau$ are mostly smooth in time, punctuated by sudden jumps whenever a photon is detected. This 
suggests the following simple numerical strategy for solving Eqs.~(\ref{eq:sme_pc}) and~(\ref{eq:tau_pceq}).

When $dN_t=0$, Eqs.~(\ref{eq:sme_pc}) and~(\ref{eq:tau_pceq}) reduce to ordinary differential equations, so $\rho$ and $\tau$ can be propagated 
using a Runge-Kutta scheme. When $dN_t = 1$, only the jump contributions are retained, and the operators are updated according to $\rho \to c\rho c^\dagger/\langle c^\dagger c\rangle_\rho - \rho$ and $\tau \to c\tau c^\dagger / \langle c^\dagger c\rangle_\rho - \tau$. At each step, an explicit normalization of $\rho$ can be imposed to ensure $\operatorname{Tr}[\rho]=1$. In our practice, we find that a first-order Runge-Kutta method already allows for efficient access to $I_\text{pc}$ near dissipative critical points for the case $\eta=1$.

\begin{figure}[t!]
\includegraphics[clip,width=8.5cm]{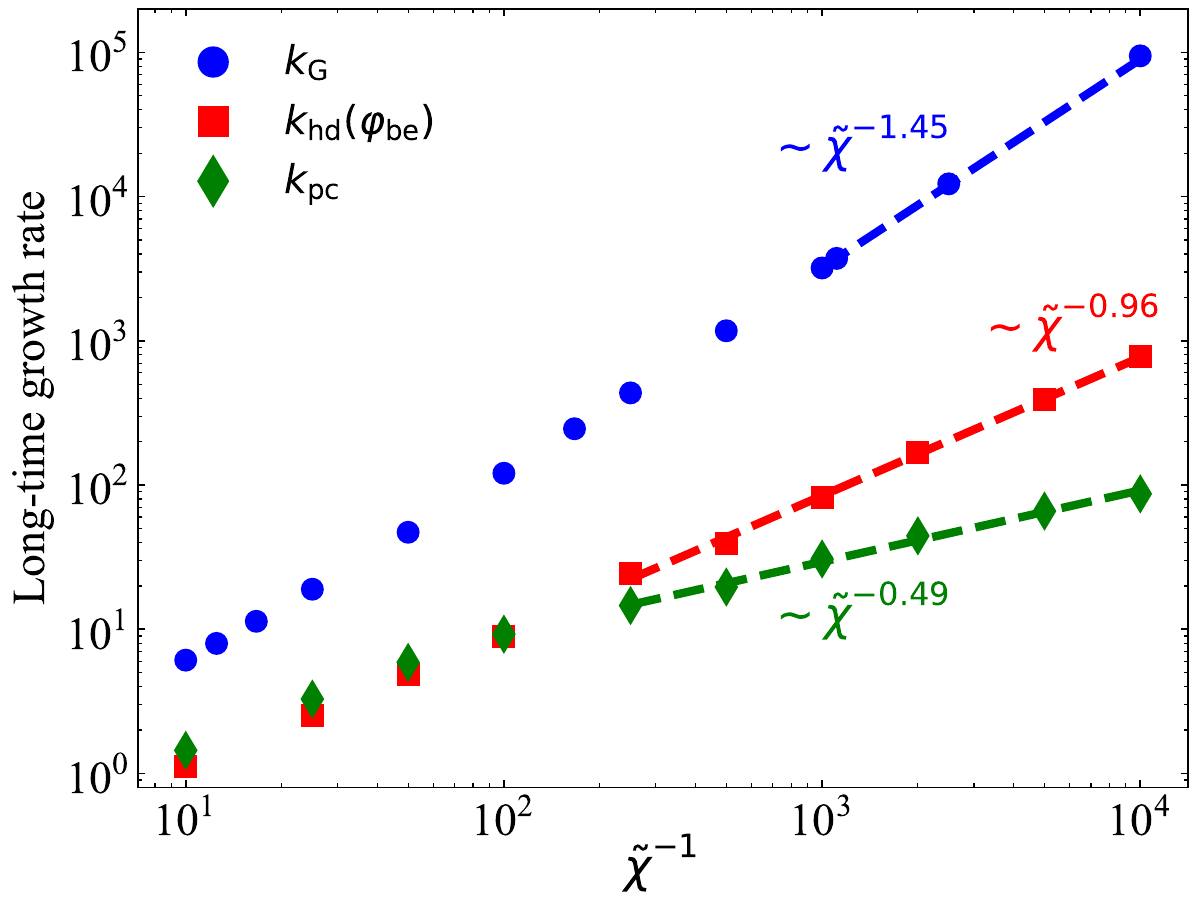}
\caption{Rates $k_\text{G}$, $k_\text{hd}(\varphi_\text{be})$ and $k_\text{pc}$ as functions of $\tilde{\chi}^{-1}$ at the dissipative critical point $\tilde{\epsilon}=\tilde{\epsilon}_c(\tilde{\omega})$ for $\tilde{\omega}=1.0$. The data for $k_\text{hd}(\varphi_\text{be})$ (squares) and $k_\text{pc}$ (diamonds) are obtained under the ideal detection condition $\eta=1$. Data points corresponding to larger values of $\tilde{\chi}^{-1}$ are fitted to extract the exponents $\gamma_\text{G} = 1.45$, $\gamma_\text{hd} = 0.96$ and $\gamma_\text{pc} = 0.49$, as demonstrated by the dashed lines. The points used for numerical fitting are those located within the lines.
}\label{fig:slopes}
\end{figure}

\section{The Kerr parametric oscillator under continuous monitoring}\label{sec:KPO}
In Sec.~\ref{sec:KPO}, we introduce the open KPO model and important known results regarding the properties of the dissipative phase transitions emerged in this setup. This model has recently attracted extensive interest in the context of quantum critical sensing~\cite{DiCandia2023,PhysRevLett.133.040801,zhang2025enhancinginformationretrievalquantumoptical}. In Sec.~\ref{sec:knownresults}, we summarize key findings for the KPO sensor from our previous work~\cite{zhang2025enhancinginformationretrievalquantumoptical}, which clarify the motivation behind the scaling analysis of the Fisher information presented in the following section.

\subsection{Sensor model}\label{sec:KPO}
The KPO model describes a Kerr oscillator with bare frequency $\omega_c$ subject to a parametric (two-photon) pump at frequency $\omega_p$. In the laboratory frame, the Kerr oscillator is described by the Hamiltonian~\cite{PhysRevLett.133.040801}
\begin{equation}\label{eq:KerrHamOrig}
H_\text{KPO} = \omega_c a^\dagger a + \frac{\epsilon}{2}\left(a^{\dagger2}e^{i\omega_p t} + a^2e^{-i\omega_p t}\right) + \chi a^{\dagger2} a^2, 
\end{equation}
where $a$ and $a^\dagger$ are the annihilation and creation operators of the oscillator, satisfying $[a,a^\dagger]=1$, where $\epsilon$ is the amplitude of the parametric pump, and where $\chi$ denotes the Kerr nonlinearity. The validity of this Hamiltonian is justified in the regime $\lvert \omega_c - \omega_p/2 \rvert \ll \omega_c$~\cite{PhysRevLett.133.040801}. 

To remove the explicit time dependence in $H_\text{KPO}$, we move to a frame rotating with frequency $\omega_p/2$. In this new frame, the Hamiltonian becomes 
\begin{equation}\label{eq:Hw}
\begin{array}{lll}
H_\omega &\displaystyle\equiv U H_\text{KPO} U^\dagger - iU\partial_t{U}^\dagger \\
&\displaystyle= \omega a^\dagger a + \frac{\epsilon}{2}(a^{\dagger2} + a^2) + \chi a^{\dagger2} a^2, 
\end{array}
\end{equation}
where $U = \exp(-i\omega_pa^\dagger a t/2)$ is the unitary implementing the rotation, and $\omega = \omega_c - \omega_p/2$ is the oscillator detuning relative to half of the pump frequency. The validity condition mentioned above implies the constraint $\lvert \omega\rvert \ll \omega_c$. In this work, we consider the estimation of the detuning $\omega$, meaning that $\theta=\omega$ and $H_\theta = H_\omega$ in Eq.~(\ref{eq:me}). To account for photon loss in the Kerr oscillator, the jump operator in the Lindblad master equation~(\ref{eq:me}) is taken as $c = \sqrt{\kappa}a$, where $\kappa$ denotes the photon loss rate. 

To simplify the notation, in what follows, we introduce the following dimensionless quantities: $\tilde{t}\equiv \kappa t$, $\tilde{\omega}\equiv\omega/\kappa$, $\tilde{\epsilon}\equiv \epsilon/\kappa$, and $\tilde{\chi} \equiv \chi/\kappa$. 

This model possesses a $\mathbb{Z}_2$ symmetry, as can be readily verified by the invariance of the Lindblad master equation~(\ref{eq:me}) under the transformation $a \to -a$. As a result, the steady-state $\varrho_\text{ss}$, i.e., the stationary solution to Eq.~(\ref{eq:me}), either preserves this symmetry in the normal phase or breaks it spontaneously within the symmetry-broken phase in the scaling limit $\tilde{\chi}\to0$. Interestingly, for $\tilde{\omega}>0$ the transition between these two phases is continuous, whereas, for $\tilde{\omega}<0$, it is first-order~\cite{PhysRevLett.36.1135,Drummond_1980,PhysRevA.95.012128,PhysRevA.98.042118,PhysRevA.106.033707,PhysRevA.111.L040201}. This difference arises only in dissipative phase transitions, and is absent in the closed counterpart of this model. The phase boundary for $\tilde{\omega}>0$ is parametrized, in the $\tilde{\omega}$-$\tilde{\epsilon}$ plane, as the curve $\tilde{\epsilon}=\tilde{\epsilon}_c(\tilde{\omega})$ with~\cite{DiCandia2023,PhysRevLett.133.040801}
\begin{equation}
\tilde{\epsilon}_c(\tilde{\omega}) = \sqrt{\tilde{\omega}^2+1/4}, 
\end{equation}
while the boundary for $\tilde{\omega}<0$ can only be determined numerically. At $\tilde{\omega}=0$, these phases boundaries intersect, and a tri-critical point emerges exhibiting a multicritical behavior~\cite{PhysRevA.111.L040201}.

\subsection{Long-time behaviors of the Fisher information in the regime of vanishing Kerr nonlinearity and ideal detection}\label{sec:knownresults}
In this subsection, we summarize the key findings from our earlier work~\cite{zhang2025enhancinginformationretrievalquantumoptical}, since they are directly relevant to the scaling analysis presented in the next section. In that work, we examined the long-time behavior of both the quantum and classical Fisher information along the phase boundary $\tilde{\epsilon}=\tilde{\epsilon}_c(\omega)$ in the limit of vanishing Kerr nonlinearity, $\tilde{\chi}=0$, and unit efficiency, $\eta=1$. 

First, we found that in the long-time limit $\tilde{t}\to+\infty$, the Fisher information scales as
\begin{equation}\label{eq:IGFpcscaling}
I_\text{G}\sim \tilde{t}^4,~~~I_\text{pc} \sim \tilde{t}^2, 
\end{equation}
where the first relation was derived rigorously, while the second was obtained numerically.

Second, we showed that the long-time scaling of $I_\text{hd}(\varphi)$ depends sensitively on the homodyne phase $\varphi$. Specifically, for a fixed $\tilde\omega$, the asymptotic behavior is given by
\begin{equation}\label{eq:Fhdscaling}
I_\text{hd}(\varphi) \sim
\left\{
\begin{array}{@{}ll@{}}
\tilde{t}^4,   & \text{for}~ \varphi = \varphi_\text{be}(\tilde\omega), \\[4pt]
\tilde{t}, & \text{otherwise},
\end{array}
\right.
\end{equation}
where $\varphi_\text{be}(\tilde\omega)=\tan^{-1}(1/2\tilde\omega)/2$. We refer to 
\begin{equation}\label{eq:becondition}
\varphi = \varphi_\text{be}(\tilde\omega)
\end{equation} 
as the backaction-evading condition, for reasons that will become clear shortly.

Equations~(\ref{eq:IGFpcscaling}) and (\ref{eq:Fhdscaling}) clearly demonstrate the advantage of continuous homodyne detection over continuous photon counting in the regime $\tilde{\chi}=0$. In fact, for a given $\tilde{\omega}>0$ to be estimated, the optimal continuous homodyne measurement is achieved when $\varphi = \varphi_\text{be}(\tilde\omega)$, in which case the asymptotic behavior of $I_\text{hd}(\varphi_\text{be})$ scales identically as $I_\text{G}$, i.e., $I_\text{G}, I_\text{hd}(\varphi_\text{be}) \sim \tilde{t}^4$. Importantly, in Ref.~\cite{zhang2025enhancinginformationretrievalquantumoptical}, we showed that this condition is a direct consequence of the well-known mechanism of backaction evasion in quantum physics~\cite{Scully}, which operates here in the context of a dissipative phase transition. This intuitive explanation also justifies our nomenclature choice.

However, Ref.~\cite{zhang2025enhancinginformationretrievalquantumoptical} focused exclusively on the regime of $\tilde{\chi}=0$, leaving open the question of whether continuous homodyne detection retains its advantage under finite Kerr nonlinearity. It is therefore necessary to extend the scaling analysis of the Fisher information to the dissipative critical points $\tilde{\epsilon}=\tilde{\epsilon}_c(\omega)$ for $\tilde{\chi}\neq0$. This is especially important and relevant given that a nonzero Kerr nonlinearity is required to regularize the sensor dynamics in the symmetry-broken phase, whereas the dissipative phase transition in the KPO sensor is strictly well-defined only in the limit $\tilde{\chi}\to0$. From this perspective, the Kerr nonlinearity is an essential quantum resource and understanding how the Fisher information scales with $\tilde{\chi}$ is key to clarifying the metrological advantage of backaction-evading homodyne detection.

\begin{figure*}[t!]
\includegraphics[clip,width=18cm]{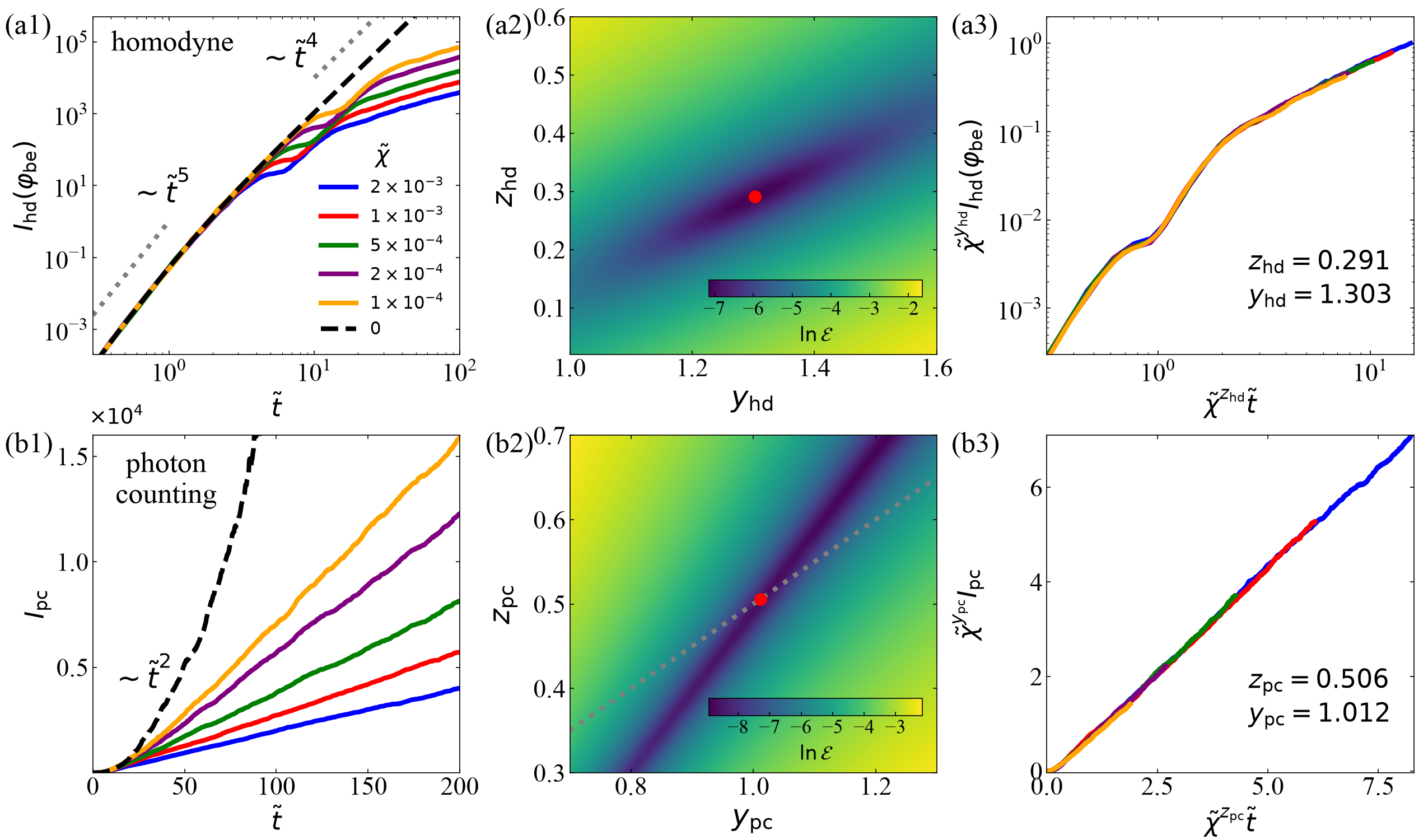}
\caption{Numerical analysis for the scaling of the classical Fisher information $I_\text{hd}(\varphi_\text{be})$  for homodyne detection in panels (a1)-(a3), and $I_\text{pc}$ for photon counting in panels (b1)-(b3). In particular, the data collapse performed in panels (a2) and (b2) is used to determine the optimal scaling exponents $y_s$ and $z_s$ for $s=\text{hd, pc}$ explicitly shown in panels (a3) and (b3). All numerical data for the classical Fisher information are obtained at the dissipative critical point $\tilde\epsilon = \tilde{\epsilon}_c(\tilde{\omega})$ with $\tilde{\omega}=1.0$, under ideal detection condition $\eta=1$. Panels (a1) and (c1) show $I_\text{hd}(\varphi_\text{be})$ and $I_\text{pc}$ as functions of time for selected finite $\tilde{\chi}$ (solid lines). The Gaussian case $\tilde{\chi}=0$ (dashed lines) is also shown, with the corresponding temporal scaling explicitly indicated (gray texts). Panels (a2) and (b2) illustrate the minimization of the rescaled collapse error $\mathcal{E}$, with the optimal values marked by red dots. In panel (b2), the optimization is performed along the constraint $y_\text{pc}=2z_\text{pc}$ (dotted line). Panels (a3) and (b3) show the optimal collapse using the explicitly stated values of $y_s$ and $z_s$ (texts). In all cases, a polynomial of degree $n=12$ is used for the collapse. 
}\label{fig:CFIfig}
\end{figure*}

\section{Scaling analysis at dissipative critical points}\label{sec:scalinganalysis}
We now analyze how the Fisher information $I_\text{G}$, $I_\text{pc}$, and $I_\text{hd}(\varphi_\text{be})$ scales with $\tilde{\chi}$ at the dissipative critical points $\tilde{\epsilon}=\tilde{\epsilon}_c(\tilde{\omega})$. As motivated more in detail in Sec.~\ref{sec:QSCM}, here we set $\eta=1$, which allows the dynamics to be described by a SSE so that the classical Fisher information can be efficiently computed.

\subsection{Finite-size scaling analysis}\label{sec:FSStheory}

Since a finite Kerr nonlinearity $\tilde{\chi}$ changes the long-time behavior of the Fisher information at the dissipative critical points $\tilde{\epsilon}=\tilde{\epsilon}_c(\tilde{\omega})$, it constitutes a relevant perturbation to the Fisher information. We thus introduce the  effective time scales $\xi_{s,\tilde{\chi}} = 1/\tilde{\chi}$, with $s\in\{\text{G, hd, pc}\}$, corresponding respectively to the global, homodyne, and photon-counting Fisher information.  Following the theory of finite-size scaling~\cite{10.1093/oso/9780198517962.001.0001,ARDOUREL202399}, $I_s$ obeys the standard scaling form $I_s = \xi_{s,\tilde{\chi}}^{y_s} \mathcal{I}_s(\xi_{s,\tilde{\chi}}^{z_s} \tilde{t})$, where $\mathcal{I}_s(x)$ are universal scaling functions independent of all microscopic parameters of the KPO sensor, and $y_s,\,z_s>0$ are two scaling exponents to be determined. Substituting $\xi_{s,\tilde{\chi}}$ into this form yields
\begin{equation}\label{eq:Iansatz}
I_s = \frac{1}{\tilde{\chi}^{y_s}} \mathcal{I}_s \left(\tilde{\chi}^{z_s}\tilde{t}\right). 
\end{equation}
This ansatz must reproduce two well-defined limiting behaviors. 

(i) In the short-time regime $\tilde{t} \ll 1/\tilde{\chi}^{z_s}$, $I_s$ follows a power law in time, $I_s = c_0^s\tilde{t}^{\alpha_s}$, where $c_0^s$ are constants independent of $\tilde\chi$ and where $\alpha_s>0$ are the scaling exponents governing the temporal behavior of $I_s$ for $\tilde{\chi}=0$. To ensure that the ansatz in Eq.~(\ref{eq:Iansatz}) reproduces this form, the scaling functions must behave as $\mathcal{I}_s(x)= c_0^s x^{\alpha_s}$ for $x\ll1$, leading to the relation 
\begin{equation}\label{eq:yI}
y_s = \alpha_s z_s, 
\end{equation}
which ensures the vanishing of $\tilde{\chi}$ in this regime.

(ii) In the long-time regime $\tilde{t} \gg 1/\tilde{\chi}^{z_s}$, the quantities $I_s$ grow linearly with time, $I_s= k_s\tilde{t}$. The ansatz in Eq.~(\ref{eq:Iansatz}) captures this behavior only if $\mathcal{I}_s(x) = c_s^\infty x$ as $x\to+\infty$, where $c_s^\infty$ are positive constants independent of $\tilde{\chi}$. From this asymptotic form we obtain $k_s = c_s^\infty \tilde{\chi}^{-\gamma_s}$, in terms of the exponents 
\begin{equation}\label{eq:gammaI}
\gamma_s \equiv y_s - z_s = (\alpha_s-1)z_s, 
\end{equation}
which can therefore be extracted by examining how the long-time growth rates $k_s$ scale with $\tilde{\chi}^{-1}$. As discussed later,  this method enables a relatively accurate estimation of $\gamma_s$, and it serves as an independent check in addition to the standard data collapse approach that we describe in the next subsection.

\subsection{Global quantum Fisher information}
We first present a numerical analysis of the scaling behavior of the global quantum Fisher information. In Fig.~\ref{fig:QFIfig}(a) we show the numerical results for $I_\text{G}$ (symbols) at the dissipative critical point $\tilde{\epsilon}=\tilde{\epsilon}_c(\tilde{\omega})$ for $\tilde{\omega}=1.0$ and for selected  values of $\tilde\chi$, including the Gaussian case $\tilde{\chi}=0$ (dashed line). In all cases, $I_\text{G}\sim\tilde{t}^4$ at short times, before crossing over to a linear growth $I_\text{G}\sim\tilde{t}$ on a time scale that increases as $\tilde{\chi}$ is reduced. The short-time behavior $I_\text{G}\sim\tilde{t}^4$ is consistent with both the numerical result for $\tilde{\chi}=0$ (dashed line) and the analytical prediction in Eq.~(\ref{eq:IGFpcscaling})~\cite{zhang2025enhancinginformationretrievalquantumoptical}.  

To extract the exponents $y_\text{G}$ and $z_\text{G}$, we collapse the data shown in Fig.~\ref{fig:QFIfig}(a), denoted collectively by the set $D= \{(\tilde{t}_i,d_i)\vert i=1,2,\cdots\}$, using the ansatz in Eq.~(\ref{eq:Iansatz}). Before performing the collapse, the data are rescaled according to $\bar{t}_i = \tilde{\chi}^{z_\text{G}}\tilde{t}_i$ and $\bar{d}_i = \tilde{\chi}^{y_\text{G}}d_i$, which are then fitted in the range $\operatorname{min}_i\bar{t}_i\le t \le \operatorname{max}_i\bar{t}_i$ with a polynomial of degree $n$, 
\begin{equation}
\mathtt{p}(t) = \sum_{k=1}^n c_kt^n, 
\end{equation}
where $c_k\in\mathbb{R}$ are fitting parameters. Since $I_\text{G}$ vanishes at the initial time $\tilde{t}=0$, i.e., $I_\text{G}(\tilde{t}=0)=0$, we consistently set the constant term in $\mathtt{p}(t)$  
to zero so that $\mathtt{p}(0)=0$.

To assess the quality of the collapse, we compute the scaled error
\begin{equation}
\mathcal{E} = \frac{ \operatorname{Mean}\bigl[(\bar{d}_i - \hat{\bar{d}}_i)^2\bigr] }{ \operatorname{Var}[\bar{d}_i] } , 
\end{equation}
where $\hat{\bar{d}}_i = \mathtt{p}(\bar{t}_i)$ are the fitted values at the rescaled time $\bar{t}_i$, where $\operatorname{Mean}[(\bar{d}_i - \hat{\bar{d}}_i)^2]$ denotes the arithmetic average of the squared residuals, and where $\operatorname{Var}[\bar{d}_i]$ is the variance of the rescaled data. Dividing by the variance ensures that this error measure remains meaningful even when the rescaled data span significantly different orders of magnitude.

The optimal data collapse is obtained by minimizing $\mathcal{E}$ over the exponents $y_\text{G}$ and $z_\text{G}$, i.e., by computing 
\begin{equation}
\operatorname{argmin}_{y_\text{G}, z_\text{G}}\,\mathcal{E}.
\end{equation}
This optimization procedure is illustrated in Fig.~\ref{fig:QFIfig}(b) with a polynomial of degree $n=12$, yielding the optimal values $y_\text{G}=1.988 $ and $z_\text{G} = 0.53$ (marked by the red dot). Using the relations in Eqs.~(\ref{eq:yI}) and (\ref{eq:gammaI}), this further implies $\alpha_\text{G} = 3.75$ and $\gamma_\text{G} = 1.458$. The corresponding optimal collapse is demonstrated in the main plot of Fig.~\ref{fig:QFIfig}(c).

These values are fully consistent with the theoretical predictions of Ref.~\cite{PRXQuantum.3.010354}, which, in the notation of this work, can be expressed as 
\begin{equation}\label{eq:IGtheory}
I_\text{G} \sim
\left\{
\begin{array}{@{}ll@{}}
\tilde{t}^{-2\Delta_{a^\dagger a}/z+2}, & \text{for}~\tilde{t}\lesssim \tilde{\chi}^z, \\[4pt]
\tilde{t}\tilde{\chi}^{-2\Delta_{a^\dagger a}+z}, & \text{for}~\tilde{t}\gg \tilde{\chi}^z.
\end{array}
\right.
\end{equation}
Here, $z$ is the dynamical exponent defined through the scaling of the Liouvillian gap, $\delta_\mathcal{L} \sim \tilde{\chi}^z$, and $\Delta_{a^\dagger a}$ is the scaling dimension of the photon-number operator $a^\dagger a$, defined by $\langle a^\dagger a\rangle_{\varrho_\text{ss}} \sim \tilde{\chi}^{\Delta_{a^\dagger a}}$, where $\langle a^\dagger a\rangle_\text{ss} = \operatorname{Tr}[a^\dagger a\varrho_\text{ss}]$ denotes the steady-state photon number. The number operator $a^\dagger a$ is relevant here because it encodes the parameter $\omega$, see the definition of the Kerr Hamiltonian in Eq.~(\ref{eq:Hw}). Both $\delta_\mathcal{L}$ and $\langle a^\dagger a\rangle_\text{ss}$ are evaluated at the dissipative critical points $\tilde{\epsilon}=\tilde{\epsilon}_c(\tilde{\omega})$. Note that this expression extends the general result of Ref.~\cite{PRXQuantum.3.010354}, which was originally derived for quantum many-body systems with spatial dimension $d\ge1$. Equation~(\ref{eq:IGtheory}) is compatible with the ansatz in Eq.~(\ref{eq:Iansatz}) since it predicts $\alpha_\text{G} = -2\Delta_{a^\dagger a}/z+2$ and $\gamma_\text{G} = -2\Delta_{a^\dagger a}+z$, which exactly obey the relation in Eq.~(\ref{eq:gammaI}), provided the identification $z_\text{G} = z$.

The exponents $z$ and $\Delta_{a^\dagger a}$ are determined in the inset of Fig.~\ref{fig:QFIfig}(c), yielding $z=0.5$ and $\Delta_{a^\dagger a}=-0.5$. Substituting these values into Eq.~(\ref{eq:IGtheory}) gives $\alpha_\text{G} = 4$ and $\gamma_\text{G} = 1.5$, in full agreement with the values obtained from the data collapse.

As an independent check, we also extract $\gamma_\text{G}$ by fitting the long-time behavior of $I_\text{G}$ linearly for $\tilde{\chi}^{-1}$ ranging from $10^1$ to $10^4$. This yields the growth rate $k_\text{G}$ as a function of $\tilde{\chi}$, which we show in Fig.~\ref{fig:slopes}. Fitting the data points corresponding to the larger values of $\tilde{\chi}$ gives $k_\text{G} \sim \tilde{\chi}^{-1.45}$, and hence $\gamma_\text{G}=1.45$, consistent with the previous value obtained from the collapse procedure. In Table~\ref{tab:exponents}, we summarize the analytical relations in Eq.~(\ref{eq:yI}) and  Eq.~(\ref{eq:gammaI}), and present the list of the theoretical  critical exponents, estimated as a rational approximation of the corresponding numerical values.

We note that the scaling with respect to $\tilde{\chi}$ directly provides also the scaling with respect to other metrological resources, such as the steady-state photon number $\langle a^\dagger a\rangle_{\varrho_\text{ss}}$. To establish this connection, one inverts the relation $\langle a^\dagger a\rangle_{\varrho_\text{ss}} \sim \tilde{\chi}^{\Delta_{a^\dagger a}}$ and substitutes the resulting expression for $\tilde{\chi}$ into $k_\text{G} = c_\infty\tilde{\chi}^{-\gamma_\text{G}}$, obtaining $k_\text{G} \sim \langle a^\dagger a\rangle_{\varrho_\text{ss}}^3$. We note that this value exceeds the standard quantum limit $\sim \langle a^\dagger a\rangle_{\varrho_\text{ss}}^1$.

\begin{table}[t]
\setlength{\tabcolsep}{10pt}
\centering
\begin{tabular}{lccccc}
\toprule
 & $z_s$ & $\alpha_s$ & $y_s = \alpha_s z_s$ & $\gamma_s=y_s-z_s$  \\
\midrule
$I_\text{G}$  & $1/2$ & $4$ & $2$ & $3/2$  \\
$I_\text{hd}(\varphi_\text{be})$  & $1/3$ & $4$ & $4/3$ & $1$  \\
$I_\text{pc}$  & $1/2$ & $2$ & $1$ & $1/2$  \\
\bottomrule
\end{tabular}
\caption{Inferred theoretical values of the critical exponents governing the scaling behaviors of the Fisher information, based on finite-size scaling analysis of the numerical results presented in this work. The first row additionally displays the 
analytical relations among the critical exponents.
}\label{tab:exponents}
\end{table}

\subsection{Classical Fisher information}\label{sec:CFIscaling}
In this section, we provide the numerical analysis of the classical Fisher information for  continuous homodyne detection and continuous photon counting. In order to extract the scaling exponents, we are going to follow the same procedure described in the previous subsection, consisting in plotting the classical Fisher information as a function of the rescaled time, collapsing the data using the scaling hypothesis, and finding the exponents through a similar optimization procedure.

\subsubsection{Continuous homodyne detection}
Following the procedure highlighted above, we start by plotting the classical Fisher information for continuous homodyne detection $I_\text{hd}(\varphi_\text{be})$ in Fig.~\ref{fig:CFIfig}(a1). We then collapse the data which allows us to find the optimal values for the exponents $y_\text{hd}$ and $z_\text{hd}$ as shown in Fig.~\ref{fig:CFIfig}(a2) and (a3). 
We note that, for example, for a fixed $\tilde{\chi}$, such as 
$\tilde{\chi}=1\times10^{-4}$, the classical Fisher information for homodyne detection $I_\text{hd}(\varphi_\text{be})$ is two orders of magnitude smaller than the value for the global quantum Fisher information $I_\text{G}$ at large times. As a consequence, we conclude that backaction-evading homodyne detection is far from optimal. 

In general, $I_\text{hd}(\varphi_\text{be})$ grows as a power-law at short times, and it correctly converges to the behavior for the Gaussian model as $\tilde{\chi}\to0$. In this $\tilde{\chi}=0$ regime, the power-law asymptotics slows down from $\tilde{t}^5$ to $\tilde{t}^4$ around $\tilde{t}=10$, and the exponent corresponding to the latter case coincides with the analytical prediction in Eq.~(\ref{eq:IGFpcscaling})~\cite{zhang2025enhancinginformationretrievalquantumoptical}. This slowdown has a profound influence on the data collapse, as we will describe in more detail below.

Interestingly, for the nonlinear model $\tilde{\chi}\neq0$, the Fisher information $I_\text{hd}(\varphi_\text{be})$  
exhibits a further characteristic oscillation before eventually crossing over to a linear asymptotics. This behavior facilitates the collapse of $I_\text{hd}(\varphi_\text{be})$ and enables a more precise determination of the exponents $y_\text{hd}$ and $z_\text{hd}$, as reflected by the semi-major axis of the error ellipse [darker region in Fig.~\ref{fig:CFIfig}(a2)] which is shorter with respect to all other cases. Note that, to collapse both the oscillatory and linear regimes simultaneously, the logarithm of the rescaled data must be used. The optimal exponents are found to be $z_\text{hd}=0.291$ and $y_\text{hd}=1.303$, and the corresponding optimal collapse of $I_\text{hd}(\varphi_\text{be})$ is shown in Fig.~\ref{fig:CFIfig}(a3). 

Using the relations in Eqs.~(\ref{eq:yI}) and (\ref{eq:gammaI}), these results also provides the values for the exponents  
$\gamma_\text{hd} = 1.012$ and $\alpha_\text{hd} = 4.48$. The former is consistent with the result $\gamma_\text{hd} = 0.96$ obtained from a linear numerical fit of $k_\text{hd}(\varphi_\text{be})$, as shown in Fig.~\ref{fig:slopes}. The latter value, however, lies between $4$ and $5$, corresponding respectively to the long-time behavior $I_\text{hd}(\varphi_\text{be})\sim \tilde{t}^4$ and the short-time behavior $I_\text{hd}(\varphi_\text{be})\sim \tilde{t}^5$ for $\tilde{\chi}=0$, as indicated by the dotted lines in Fig.~\ref{fig:CFIfig}(a1). This may be attributed to the fact that the oscillatory behaviors of the curves in Fig.~\ref{fig:CFIfig}(a1) lie between these two regimes. It is therefore reasonable to expect that the estimation of $\alpha_\text{hd}$ would approach $4$ if smaller values of $\tilde{\chi}$ were included in the collapse. 

The results for $\gamma_\text{hd}$ are sufficiently accurate to support the theoretical prediction $\gamma_\text{hd}=1$. 
If the considerations above about the oscillatory behavior of the curves in Fig.~\ref{fig:CFIfig}(a1) are correct, so that 
$\alpha_\text{hd}=4$, then the theoretical values for $y_\text{hd}$ and $z_\text{hd}$ are $y_\text{hd} = 4/3$ and $z_\text{hd} = 1/3$, which are compatible with the numerical results obtained from data collapse. From these values, the photon-number scaling follows directly as $k_\text{hd}(\varphi_\text{be}) \sim \langle a^\dagger a\rangle_{\varrho_\text{ss}}^2$, which exceeds the standard quantum limit even though it does not saturate the exponent found for $I_\text{G}$.

\subsubsection{Continuous photon counting}
Finally, we analyze the scaling behavior of the classical Fisher information for photon counting. As in the other cases, we proceed to determine the exponents $y_\text{pc}$ and $z_\text{pc}$ using the numerical data shown in Fig.~\ref{fig:CFIfig}(b1) for the same values of $\tilde{\chi}$, which are collapsed in Fig.~\ref{fig:CFIfig}(b2) to find the optimal estimate. 
In contrast to the homodyne case, this procedure does not 
fully converge as the collapse error $\mathcal{E}$ remains small in an extended parameter region, as demonstrated by the dark stripe in Fig.~\ref{fig:CFIfig}(b2). To circumvent this issue, we instead minimize $\mathcal{E}$ under the constraint $y_\text{pc}=2z_\text{pc}$ (gray dotted line), which is derived from Eq.~(\ref{eq:yI}) using $\alpha_\text{pc}=2$. Here we use the value $\alpha_\text{pc}=2$ because by definition the exponent $\alpha_\text{pc}$ governs the temporal scaling of $I_\text{pc}$ for $\tilde{\chi}=0$, i.e., $I_\text{pc}\sim \tilde{t}^{\alpha_\text{pc}}$. The result in Eq.~(\ref{eq:IGFpcscaling}), which is also verified numerically in Fig.~\ref{fig:CFIfig}(b1), thereby justifies this choice. Such a constrained collapse allows the optimization to converge to the values  
$z_\text{pc}=0.506$ and $y_\text{pc}=1.012$, as shown in Fig.~\ref{fig:CFIfig}(b2). The feasibility of the collapse procedure can be checked by the matching of the curves in Fig.~\ref{fig:CFIfig}(c1).

Using these results in Eqs.~(\ref{eq:yI}) and (\ref{eq:gammaI}), one obtains 
$\gamma_\text{hd} = 0.506$, which is also in good agreement with the result $\gamma_\text{hd}=0.49$ from fitting $k_\text{pc}$, as shown in Fig.~\ref{fig:slopes}. The theoretical values for these exponents based on the numerical analysis presented here are summarized in Table~\ref{tab:exponents}. In this case, the estimate for the
photon-number scaling is $k_\text{pc} \sim \langle a^\dagger a\rangle_{\varrho_\text{ss}}^1$, which is exactly the standard quantum limit.

\section{Conclusions and discussion}\label{sec:conclusions}
We have analyzed the scaling behavior of the Fisher information with respect to the Kerr nonlinearity at the dissipative critical points of the KPO model. In particular, our results focus on both the methodology and the physics of continuous-measurement-based critical quantum sensing. 

On the methodological side, we have extended existing methods~\cite{PhysRevA.91.012118,Albarelli2018restoringheisenberg} for computing the classical Fisher information of continuous homodyne detection. In particular, we have implemented and benchmarked higher-order time-discrete approximations with improved convergence properties. These developments enable efficient numerical access to the classical Fisher information in regimes where existing methods become unreliable, and they serve as the methodological basis for the numerical results presented in this work.

On the physical side, we have performed a finite-size scaling analysis of the Fisher information at the dissipative critical points of the KPO. The extracted critical exponents, summarized in Table~\ref{tab:exponents}, reveal a clear hierarchy among the three quantities considered. Specifically, the global quantum Fisher information $I_\text{G}$ exhibits the most favorable scaling, with a Kerr-nonlinearity exponent $\gamma_\text{G}=3/2$ and an asymptotic behavior $I_\text{G}\sim\tilde{t}^4$ in time. The scaling for continuous photon counting is markedly weaker, with $\gamma_\text{pc}=1/2$ and $I_\text{pc}\sim\tilde{t}^2$. On the other hand, continuous homodyne detection under the backaction-evading condition~(\ref{eq:becondition}) occupies an intermediate position: it achieves the same $\tilde{t}^4$ asymptotic behaviour in time as $I_\text{G}$ but with a less favorable exponent $\gamma_\text{hd}=1$.

These results have direct implications for implementing critical sensing based on the KPO. In fact, although backaction-evading homodyne detection does not saturate the quantum Cram\'er-Rao bound, it substantially outperforms photon counting, both in temporal scaling and in the scaling with respect to the Kerr nonlinearity. Moreover, by expressing the precision in terms of the steady-state photon number, we showed that backaction-evading homodyne detection attains a scaling beyond the standard quantum limit, while photon counting remains below this bound. This establishes backaction evasion as a promising strategy for enhancing precision in continuously monitored critical quantum sensors, beyond the idealized limit of vanishing Kerr nonlinearity considered in earlier work~\cite{zhang2025enhancinginformationretrievalquantumoptical}. 

Our results also leave a few open questions that deserve further investigation. First, whether backaction-evading homodyne detection represents the optimal time-local continuous measurement strategy remains an open question. Second, extending the present analysis to non-ideal detection efficiency, multi-parameter estimation~\cite{zhang2025enhancinginformationretrievalquantumoptical}, or other driven-dissipative critical systems~\cite{PhysRevLett.99.050402,PhysRevLett.118.123602,PhysRevLett.120.183603,Kirton2019,PhysRevA.103.013306,PhysRevB.107.104201} could further clarify the 
applicability and potential limitations of the backaction evasion strategy identified in this work. In particular, recent progress~\cite{ltfw-4fwn} enables the calculation of the quantum Fisher information under non-ideal detection efficiency. It is thus interesting to explore whether backaction-evading homodyne detection saturates the tighter bound set by this noisy quantum Fisher information. Lastly, it is also worthwhile to explore the application of our results in both global and local critical sensing~\cite{zhang2026criticalityenhancedglobalfrequencysensing}.

\begin{acknowledgments}
C.Z. acknowledges support from CPSF (Grants No. 2025M1773419). M.C. acknowledges support from NSFC (Grants No. 11935012 and No. 12088101) and NSAF (Grant No. U2330401). 
\end{acknowledgments}

\appendix

\section{Time-discrete approximations for simulating It\^o stochastic differential equations}\label{sec:numapprox}
In this section, we present the details of time-discrete approximations necessary for the numerical simulation of  Eq.~(\ref{eq:eq_Theta}). We thus discretize the evolution time into bins of size $\Delta t$ and denote the time at the $n$th step as $t_n=n\Delta t$. The initial time is chosen as $t_0=0$ without loss of generality. The value of the vector $\Theta$ at time $t_n$ is denoted by $\Theta_n = \Theta(t_n)$. 

\subsection{Convergence criteria}\label{sec:criteria}

The convergence speed of time-discrete approximations can be assessed under different error criteria. In particular, if the convergence of individual paths of $\Theta$ is concerned, the error criterion is given by~\cite{Platen}
\begin{equation}
\mathbb{E}[\| \Theta(t) - \Theta_n \|] \le C_\text{s} (\Delta t)^{\gamma_\text{s}}, 
\end{equation}
where $\|\Theta\| \equiv \|\rho\| + \|\tau\|$, with $\|\rho\|$ and $\|\tau\|$ denoting the $2$-norm of the density operators $\rho$ and $\tau$ respectively, $C_\text{s}$ is a constant independent of $\Delta t$, and the exponent $\gamma_\text{s}>0$ characterizes the order of convergence, usually referred to as the strong convergence order. 

A weaker convergence condition can be applied if one is only interested in the statistical properties of the components of $\Theta$, formally defined through a complex-valued function $g(\Theta)$. As an example, $g(\Theta) = \operatorname{Tr}[\tau]$ appears in Eq.~(\ref{eq:Ftau}). In this case, the criterion is given by~\cite{Platen}
\begin{equation}
\lvert \mathbb{E}[g(\Theta(t))] - \mathbb{E}[g(\Theta_n)] \rvert \le C_\text{w} (\Delta t)^{\gamma_\text{w}}, 
\end{equation}
where $C_\text{w}$ is also a constant independent of $\Delta t$, and the order of weak convergence is measured by the exponent $\gamma_\text{w}>0$.

\subsection{Three approximation schemes}\label{sec:}
We now implement three approximation schemes with different strong and weak convergence orders. These approximations are derivative-free, in the sense that no explicit derivatives of the drift $\boldsymbol{\mathcal{A}}(\Theta)$ and diffusion $\boldsymbol{\mathcal{B}}(\Theta)$ coefficients with respect to $\Theta$ are involved. This feature makes them amenable to numerical simulations. 

\subsubsection{Milstein scheme}
A convenient approximation widely used in the literature is the Milstein scheme~\cite{Platen,Albarelli2018restoringheisenberg}, which has  convergence orders $\gamma_\text{s} = \gamma_\text{w} =1.0$. In this scheme, the iteration rule is
\begin{equation}
\begin{aligned}
\Theta_{n+1} &= \Theta_n + \boldsymbol{\mathcal A}(\Theta_n) \Delta t + \boldsymbol{\mathcal B}(\Theta_n) \Delta w \\
&~~~ + \boldsymbol{\mathcal B}\bigl(\boldsymbol{\mathcal B}(\Theta_n)\bigr)\frac{\Delta^2 w - \Delta t}{2}, 
\end{aligned}
\end{equation}
where $\Delta w$ is the discretized Wiener increment. If only the first line is retained, the scheme reduces to the Euler-Maruyama method~\cite{Platen,Albarelli2018restoringheisenberg}, which has convergence orders $\gamma_\text{s} = 0.5$ and $\gamma_\text{w} = 1.0$. The correction in the second line therefore  improves the strong convergence order from $0.5$ to $1.0$, while leaving the weak convergence order unchanged. This is because the correction term vanishes upon taking the ensemble average, $\mathbb{E}[\Delta^2 w - \Delta t] = 0$, and hence does not contribute to statistical averages. We note that this scheme can also be expressed in terms of Kraus operators, as already shown and applied in Ref.~\cite{Albarelli2018restoringheisenberg}.

\subsubsection{Explicit order-2.0 weak scheme}
An improved scheme with higher weak convergence order was first proposed by Eckhard Platen~\cite{Platen}, and is hereafter referred to as the Platen scheme. The update rule is (c.f., Chapter 15.1 in Ref.~\cite{Platen})
\begin{equation}
\begin{aligned}
\Theta_{n+1} &= \Theta_n + \left[ \boldsymbol{\mathcal A}(\Theta_n) + \boldsymbol{\mathcal A}(\tilde\Theta_n) \right]\frac{\Delta t}{2} \\
&~~~ + \left[ \boldsymbol{\mathcal B}(\Theta_n^+) + \boldsymbol{\mathcal B}(\Theta_n^-) + 2\boldsymbol{\mathcal B}(\Theta_n) \right]\frac{\Delta w}{4} \\
&~~~ + \left[ \boldsymbol{\mathcal B}(\Theta_n^+) - \boldsymbol{\mathcal B}(\Theta_n^-) \right]\frac{\Delta^2 w - \Delta t}{4\sqrt{\Delta t}}, 
\end{aligned}
\end{equation}
where we have defined three intermediate vectors
\begin{equation}
\begin{aligned}
\tilde\Theta_n &= \Theta_n + \boldsymbol{\mathcal{A}}(\Theta_n) \Delta t \pm \boldsymbol{\mathcal{B}}(\Theta_n) \Delta w, \\
\Theta_n^\pm &= \Theta_n + \boldsymbol{\mathcal{A}}(\Theta_n) \Delta t \pm \boldsymbol{\mathcal{B}}(\Theta_n) \sqrt{\Delta t}.
\end{aligned} 
\end{equation}
In each iteration, the drift coefficient $\boldsymbol{\mathcal{A}}$ is evaluated twice and the diffusion coefficient $\boldsymbol{\mathcal{B}}$ three times. Thus, this approximation generally takes more time per iteration than the Milstein scheme.

\subsubsection{Explicit order-1.5 strong scheme}
We also consider a scheme with improved strong convergence order $\gamma_\text{s}=1.5$, hereafter  referred to as the Strong-1.5 scheme. The update rule is (c.f., Chapter 11.2 in Ref.~\cite{Platen})
\begin{equation}
\begin{aligned}
\Theta_{n+1} &= \Theta_n + \boldsymbol{\mathcal B}(\Theta_n) \Delta w_1 +  \left[ \boldsymbol{\mathcal A}(\Theta_n^+) - \boldsymbol{\mathcal A}(\Theta_n^-) \right] \frac{\Delta z}{2\sqrt{\Delta t}} \\
&~~~ + \left[ \boldsymbol{\mathcal A}(\Theta_n^+) + 2\boldsymbol{\mathcal A}(\Theta_n) + \boldsymbol{\mathcal A}(\Theta_n^-) \right]\frac{\Delta t}{4} \\
&~~~ + \left[ \boldsymbol{\mathcal B}(\Theta_n^+) - \boldsymbol{\mathcal B}(\Theta_n^-) \right]\frac{\Delta^2 w_1 - \Delta t}{4\sqrt{\Delta t}} \\
&~~~ + \left[ \boldsymbol{\mathcal B}(\Theta_n^+) - 2\boldsymbol{\mathcal B}(\Theta_n) + \boldsymbol{\mathcal B}(\Theta_n^-) \right]\frac{\Delta w_1\Delta t - \Delta z}{2\Delta t} \\
&~~~ + \left[ \mathbf{\mathcal B}(\Upsilon_n^+) - \boldsymbol{\mathcal B}(\Upsilon_n^-) - \boldsymbol{B}(\Theta_n^+) + \boldsymbol{\mathcal B}(\Theta_n^-) \right] \\
&~~~~~~\times \frac{(\Delta^2w_1/3 - \Delta t)\Delta w_1}{4\Delta t}, 
\end{aligned}
\end{equation}
where the Gaussian random variable $\Delta z = (\Delta w_1 + \Delta w_2/\sqrt3)\Delta t/2$, with $\Delta w_1$ and $\Delta w_2$ two independent Wiener increments, satisfies $\mathbb{E}[\Delta^2z] = \Delta^3t/3$ and $\mathbb{E}[\Delta z\Delta w_1] = \Delta^2t/2$, and two additional intermediate vectors are introduced, 
\begin{equation}
\Upsilon_n^\pm = \Theta_n^+ \pm \boldsymbol{\mathcal B}(\Theta_n^+)\sqrt{\Delta t}. 
\end{equation}
This approximation takes more computational time per iteration than the Platen scheme, since both the drift $\boldsymbol{\mathcal A}$ and diffusion $\boldsymbol{\mathcal B}$ coefficients need to be evaluated more times.

\begin{figure}
\includegraphics[clip,width=8.4cm]{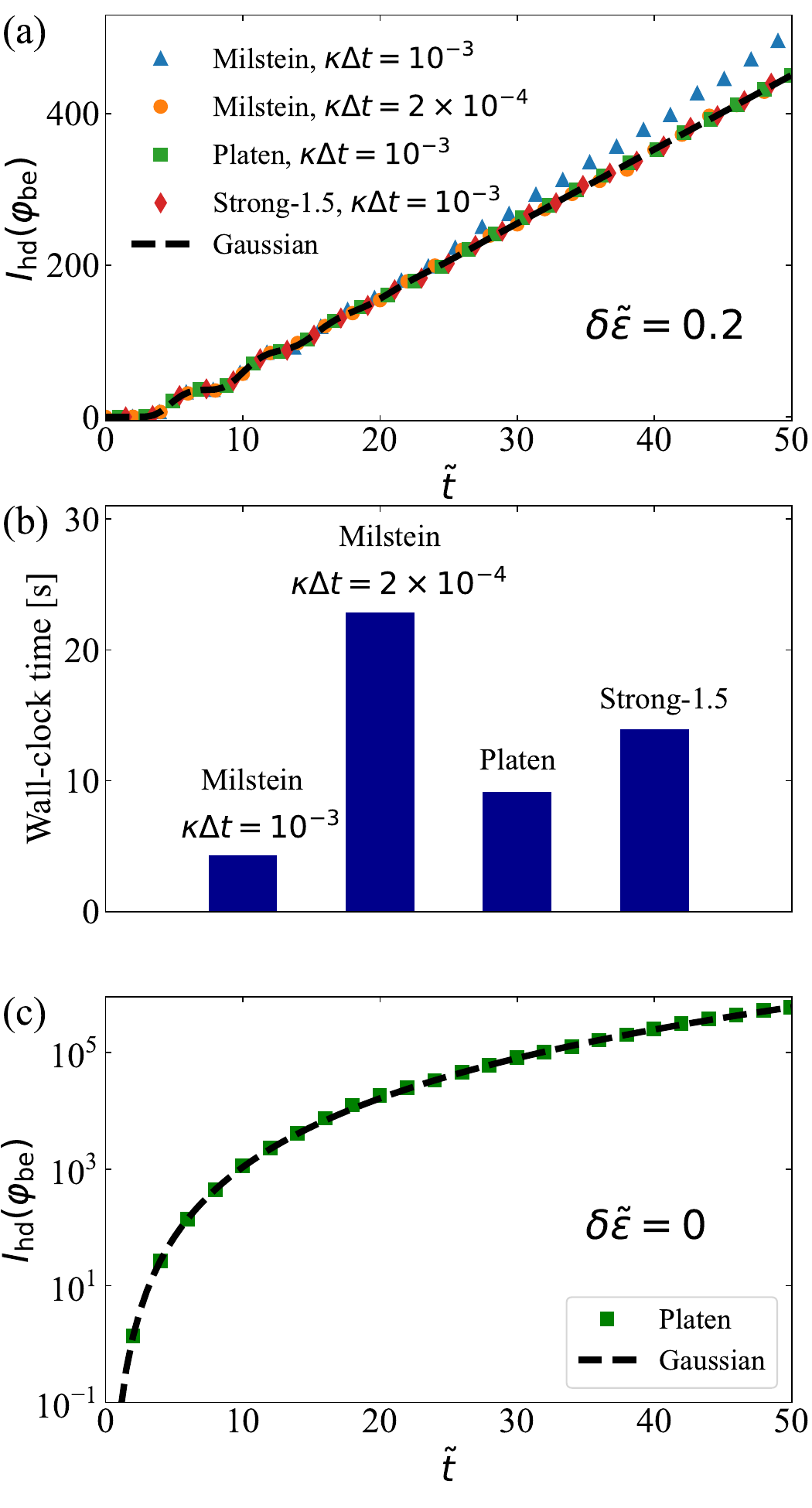}
\caption{The classical Fisher information for continuous homodyne detection, $I_\text{hd}(\varphi_\text{be})$, in the limiting regime $\tilde{\chi}=0$ and in the case of ideal detection $\eta=1$ for (a) $\delta\tilde{\epsilon} = 0.2$ and (c) $\delta\tilde{\epsilon} = 0$. The detuning is fixed at $\tilde{\omega}=1.0$. Symbols represent the results obtained from different time discrete approximations, while dashed lines correspond to the Gaussian approach detailed in Appendix~\ref{sec:app_gaussian}. In panel (a), $3600$ trajectories are simulated for each approximation scheme with the Fock-state truncation set to $N_\text{fock}=60$, whereas in panel (c) $5400$ trajectories are used for ensemble average with $N_\text{fock}=600$ and $\kappa\Delta t=5\times10^{-5}$. (b) Average wall-clock time cost per trajectory of $\Theta$ for the results shown in panel (a). 
}\label{fig:test}
\end{figure}

\subsection{Approximation schemes for ideal detection}\label{sec:sse_hd}
As noted in the main text, it is not necessary to work with the matrices $\rho$ and $\tau$ in the case of ideal detection $\eta=1$ due to the presence of the relations in Eq.~(\ref{eq:rhotau_pure}). In this case,  the quantity $\operatorname{Tr}[\tau]$ can be simplified to~\cite{Albarelli2018restoringheisenberg}
\begin{equation}\label{eq:tau_ideal}
\operatorname{Tr}[\tau] = 2\operatorname{Re}\langle\psi|\phi\rangle. 
\end{equation}

The state vector $|\psi\rangle$ obeys the SSE~\cite{Petruccione,Gardiner,Wiseman_Milburn_2009}
\begin{equation}\label{eq:sse_hd}
d|\psi\rangle = A(\psi) dt + B(\psi) dw,
\end{equation}
where the operators $A(\psi)$ and $B(\psi)$ are defined as
\begin{equation}
\begin{aligned}
A(\psi) &= \left( -iH - \frac{c^\dagger c}{2}  + \frac{\langle c_\varphi+c^\dagger_\varphi\rangle_{\psi} c}{2} - \frac{\langle c_\varphi+c^\dagger_\varphi\rangle_{\psi}^2}{8} \right) |\psi\rangle,  \\
B(\psi) &= \left(c - \frac12\langle c_\varphi+c^\dagger_\varphi\rangle_{\psi}\right) |\psi\rangle. 
\end{aligned}
\end{equation}
Both $A(\psi)$ and $B(\psi)$ are nonlinear in $\psi$ due to the presence of the terms that depend on $\langle c_\varphi+c^\dagger_\varphi\rangle_{\psi}$, which ensures the normalization of $|\psi\rangle$ to first order in $dt$. 

It has been shown that the auxiliary state vector $|\phi\rangle$ in Eq.~(\ref{eq:tau_ideal}) evolves according to the equation~\cite{Albarelli2018restoringheisenberg}
\begin{equation}\label{eq:phi}
d|\phi\rangle = -i\partial_\omega H_\omega|\psi\rangle + A(\phi) dt + B(\phi) dw. 
\end{equation}
Combining Eq.~(\ref{eq:sse_hd}) and Eq.~(\ref{eq:phi}) leads to the vector form 
\begin{equation}\label{eq:dPsi}
d |\Psi\rangle = \mathbf{A}(\Psi) dt + \mathbf{B}(\Psi) dw, 
\end{equation}
where the operators $\mathbf{A}$ and $\mathbf{B}$ are defined as
\begin{equation}
\mathbf{A}(\Psi) = \begin{pmatrix}
A(\psi) \\
-i\partial_\omega H_\omega|\psi\rangle + A(\phi)
\end{pmatrix},
\mathbf{B}(\Psi) = 
\begin{pmatrix} 
B(\psi) \\
B(\phi) 
\end{pmatrix}. 
\end{equation}
Below we present explicitly the constructions of the three approximation schemes for Eq.~(\ref{eq:dPsi}).

\subsubsection{Milstein scheme}
The update rule for the Milstein scheme is given by
\begin{equation}
\begin{aligned}
|\Psi_{n+1}\rangle &= |\Psi_n\rangle + \mathbf{A}(\Psi_n) \Delta t + \mathbf{B}(\Psi_n) \Delta w \\
&~~~+ \mathbf{B}(\mathbf{B}(\Psi_n))\frac{\Delta^2 w - \Delta t}{2}. 
\end{aligned}
\end{equation}

\subsubsection{Explicit order-2.0 weak scheme}
The update rule for the Platen scheme is given by
\begin{equation}
\begin{aligned}
|\Psi_{n+1}\rangle &= |\Psi_n\rangle +  \left[ \mathbf{A}(\Psi_n) + \mathbf{A}(\tilde\Psi_n) \right] \frac{\Delta t}{2} \\
&~~~+ \left[ \mathbf{B}(\Psi_n^+) + \mathbf{B}(\Psi_n^-) + 2\mathbf{B}(\Psi_n) \right]\frac{\Delta w}{4} \\
&~~~+ \left[ \mathbf{B}(\Psi_n^+) - \mathbf{B}(\Psi_n^-) \right]\frac{\Delta^2 w - \Delta t}{4\sqrt{\Delta t}}, 
\end{aligned}
\end{equation}
where the intermediate state vectors are
\begin{equation}
\begin{aligned}
|\tilde\Psi_n\rangle &= |\Psi_n\rangle + \mathbf{A}(\Psi_n) \Delta t + \mathbf{B}(\Psi_n) \Delta w, \\
|\Psi_n^\pm\rangle &= |\Psi_n\rangle + \mathbf{A}(\Psi_n) \Delta t \pm \mathbf{B}(\Psi_n) \sqrt{\Delta t}. 
\end{aligned}
\end{equation}

\subsubsection{Explicit order-1.5 strong scheme}
The update rule for the strong-1.5 scheme is given by
\begin{equation}
\begin{aligned}
|\Psi_{n+1}\rangle &= |\Psi_n\rangle + \mathbf{B}(\Psi_n) \Delta w_1 +  \left[ \mathbf{A}(\Psi_n^+) - \mathbf{A}(\Psi_n^-) \right] \frac{\Delta z}{2\sqrt{\Delta t}} \\
&\displaystyle+ \left[ \mathbf{A}(\Psi_n^+) + 2\mathbf{A}(\Psi_n) + \mathbf{A}(\Psi_n^-) \right]\frac{\Delta t}{4} \\
& + \left[ \mathbf{B}(\Psi_n^+) - \mathbf{B}(\Psi_n^-) \right]\frac{\Delta^2 w_1 - \Delta t}{4\sqrt{\Delta t}} \\
&+ \left[ \mathbf{B}(\Phi_n^+) - \mathbf{B}(\Phi_n^-) - \mathbf{B}(\Psi_n^+) + \mathbf{B}(\Psi_n^-) \right] \\
&~~~\times \frac{(\Delta^2w_1/3 - \Delta t)\Delta w_1}{4\Delta t} \\
& + \left[ \mathbf{B}(\Psi_n^+) - 2\mathbf{B}(\Psi_n) + \mathbf{B}(\Psi_n^-) \right]\frac{\Delta w_1\Delta t - \Delta z}{2\Delta t}, 
\end{aligned}
\end{equation}
with 
\begin{equation}
|\Phi_n^\pm\rangle = |\Psi_n^+\rangle \pm \mathbf{B}(\Psi_n^+)\sqrt{\Delta t}. 
\end{equation}

\subsection{Numerical tests}\label{sec:numtests}
Here we evaluate the numerical efficiency and accuracy of the above approximation schemes in the context of the KPO model described in Sec.~\ref{sec:KPO}, with particular focus on the regime near the dissipative critical point $\tilde\epsilon=\tilde\epsilon_c(\tilde\omega)$. These benchmarks serve both to determine the most suitable approximation for the numerical calculation of $I_\text{hd}(\varphi)$ in the vicinity of dissipative critical points and to clarify the motivations behind the approximation schemes presented earlier by numerical exemplifications.

In continuous homodyne detection, the probability distribution $p(dy_t)$ of the measured current $dy_t$ at time $t$ is Gaussian, with mean $\sqrt{\eta\kappa}\langle a e^{-i\varphi}+a^\dagger e^{i\varphi}\rangle_\rho\,dt$ and variance $dt$. It is shown in Ref.~\cite{PhysRevA.95.012116} that this important property enables the compact expression,
\begin{equation}\label{eq:Fhdsimple}
I_\text{hd}(\varphi) = 2\eta\kappa \int_0^t ds\,\mathbb{E}[(\partial_\omega \langle x_\varphi\rangle_\rho)^2], 
\end{equation}
where $x_\varphi = x\cos\varphi + p\sin\varphi$ with $x=(a+a^\dagger)/\sqrt2$ and $p=i(a^\dagger-a)/\sqrt2$ denoting the standard quadrature operators. 

Although Eq.~(\ref{eq:Fhdsimple}) is formally concise, it is not directly friedly to efficient numerical evaluation, since it requires the derivative of a stochastic quantity, $\partial_\omega \langle x_\varphi\rangle_\rho$, which is difficult to access  accurately in practice. However, this difficulty is substantially alleviated in the Gaussian limit $
\tilde{\chi}\to0$, where the solution $\rho(t)$ to the SME~(\ref{eq:sme_hd}) remains Gaussian for all times $t\ge0$, provided the initial state $\rho(0)$ is Gaussian. Details for the numerical evaluation of  $I_\text{hd}(\varphi)$ in this limit are provided in Appendix~\ref{sec:app_gaussian} and in Ref.~\cite{zhang2025enhancinginformationretrievalquantumoptical}.

In Fig.~\ref{fig:test}(a) we show the classical Fisher information $I_\text{hd}(\varphi)$ in the off-critical case for ideal detection $\eta=1$, with a proximity to the critical point of $\delta\tilde{\epsilon}=0.2$ where $\delta\tilde{\epsilon} \equiv \tilde{\epsilon}_c(\tilde\omega) - \tilde\epsilon$. The homodyne phase is fixed at the backaction-evading value $\varphi=\varphi_\text{be}(\tilde\omega)$ for $\tilde{\omega}=1.0$. The symbols correspond to the three approximation schemes introduced in Sec.~\ref{sec:sse_hd}, benchmarked against the Gaussian approach (dashed line) detailed in Appendix~\ref{sec:app_gaussian}. 

For the Platen and Strong-1.5 schemes, convergence is reached using a time step of $\kappa\Delta t = 10^{-3}$ and a Fock-state truncation of $N_\text{fock}=60$. In contrast, the Milstein scheme fails to converge under these parameters and requires a reduction of the time step to $\kappa\Delta t=2\times10^{-4}$, as demonstrated by the triangles and dots in Fig.~\ref{fig:test}. To compare numerical efficiency, we show in Fig.~\ref{fig:test}(b) the average wall-clock time cost per trajectory of $\Theta$ for each scheme. This analysis reveals that the Platen scheme demands the least computational time to achieve convergence. Moreover, in Fig.~\ref{fig:test}(c) we demonstrate that the Platen scheme also yields converging results for $I_\text{hd}(\varphi_\text{be})$ at the same critical point, while the Euler-Maruyama and the Milstein schemes fail to reach convergence under the same simulation parameters (data not shown).

Taken together, these results identify the Platen scheme as the most suitable method for computing $I_\text{hd}(\varphi)$ near dissipative critical points in the case $\eta=1$, offering the best trade-off among algorithm availability, numerical efficiency, and accuracy.

\section{Calculation of the classical Fisher information for continuous homodyne detection in the Gaussian limit $\tilde{\chi}\to0$}\label{sec:app_gaussian}
In the regime of vanishing Kerr nonlinearity, $\tilde{\chi}=0$, the open KPO model subject to continuous homodyne detection constitutes a linear Gaussian system, meaning that for any Gaussian initial state $\rho(0)$ the solution $\rho(t)$ to the SME~(\ref{eq:sme_hd}) remains Gaussian for all times $t\ge0$. Due to the preservation of its Gaussian character, $\rho(t)$ can be fully characterized by its mean vector $\mathbf{r}$ and covariance matrix $\boldsymbol{\sigma}$, defined as
\begin{equation}
\begin{aligned}
\mathbf{r} &= \begin{pmatrix}
\langle x\rangle_\rho \\
\langle p\rangle_\rho
\end{pmatrix}, \\
\boldsymbol{\sigma} &= \begin{pmatrix}
2(\langle x^2\rangle_\rho - \langle x\rangle_\rho^2) & \langle\{x,p\}\rangle_\rho - 2\langle x\rangle_\rho\langle p\rangle_\rho \\
\langle\{x,p\}\rangle_\rho - 2\langle x\rangle_\rho\langle p\rangle_\rho & 2(\langle p^2\rangle_\rho - \langle p\rangle_\rho^2) 
\end{pmatrix}, 
\end{aligned}
\end{equation}
which evolves according to the stochastic differential equations
\begin{subequations}\label{eq:rsigmaeqs}
\begin{align}
\displaystyle d\mathbf{r} &\displaystyle= \mathbf{A} \mathbf{r}dt + \sqrt{\frac{\eta\kappa}{2}}(\boldsymbol\sigma-\mathbf{I})\mathbf{v}dw, \label{eq:rsigmaeqs1} \\
\displaystyle\frac{d\boldsymbol\sigma}{dt} &\displaystyle= \mathbf{A}\boldsymbol\sigma + \boldsymbol\sigma \mathbf{A}^\intercal + \mathbf{D} - \eta\kappa(\boldsymbol\sigma - \mathbf{I})\mathbf{B}(\boldsymbol\sigma - \mathbf{I}), \label{eq:rsigmaeqs2}
\end{align}
\end{subequations}
where we have introduced the coefficient matrices 
\begin{equation}
\mathbf{A} = \begin{pmatrix}
\displaystyle-\kappa/2 & \omega-\epsilon \\
-\omega-\epsilon & \displaystyle-\kappa/2
\end{pmatrix}, 
\mathbf{v} = \begin{pmatrix}
\cos\varphi \\
\sin\varphi
\end{pmatrix}, 
\end{equation}
$\mathbf{B} = \mathbf{v}^\intercal\mathbf{v}$, $\mathbf{D} = \operatorname{diag(\kappa,\kappa)}$ and the $2\times2$ identity matrix $\mathbf{I}$. The measured photocurrent can also be expressed in terms of these matrices as $dy_t = \sqrt{2\eta\kappa}\mathbf{v}^\intercal\mathbf{r}+dw$. Derivations of these equations have been detailed in our earlier work~\cite{zhang2025enhancinginformationretrievalquantumoptical}. 

To exploit Eq.~(\ref{eq:Fhdsimple}) for the calculation of $I_\text{hd}(\varphi)$, it is necessary to find the solution of the integrand $\mathbb{E}[(\partial_\omega \langle x_\varphi\rangle_\rho)^2]$, which can be rewritten in a more concise form as
\begin{equation}\label{eq:integrand}
\mathbb{E}[(\partial_\omega \langle x_\varphi\rangle_\rho)^2] = \operatorname{Tr}\left[\mathbf{B} \mathbb{E}\left[(\partial_\omega\mathbf{r})(\partial_\omega\mathbf{r}^\intercal)\right] \right]. 
\end{equation} 
We now show that it is possible to derive a set of deterministic differential equations to compute this integrand. 

To this end, we differentiate both sides of Eq.~(\ref{eq:rsigmaeqs}) with respect to $\omega$, yielding
\begin{subequations}\label{eq:eqs_pwrSigma}
\begin{align}
\displaystyle d(\partial_\omega\mathbf{r}) &\displaystyle= (\partial_\omega\mathbf{A}) \mathbf{r}dt + \mathbf{A}(\partial_\omega\mathbf{r})dt + \sqrt{\frac{\kappa\eta}{2}}(\partial_\omega\boldsymbol\sigma)\mathbf{v}dw \nonumber \\
&\displaystyle~~~ - \kappa\eta(\boldsymbol\sigma-\mathbf{I})\mathbf{B}(\partial_\omega\mathbf{r})dt, \label{eq:eq_pwr} \\
\displaystyle\frac{d(\partial_\omega\boldsymbol\sigma)}{dt} &\displaystyle= (\mathbf{A}+\kappa\eta \mathbf{B})(\partial_\omega\boldsymbol\sigma) + (\partial_\omega\boldsymbol\sigma)(\mathbf{A}^\intercal+\kappa\eta \mathbf{B}) \nonumber\\
&\displaystyle~~~ - \kappa\eta(\partial_\omega\boldsymbol\sigma)\mathbf{B}\boldsymbol{\sigma} - \kappa\eta\boldsymbol{\sigma}\mathbf{B}(\partial_\omega\boldsymbol{\sigma}) \nonumber\\
&\displaystyle~~~ + (\partial_\omega \mathbf{A})\boldsymbol\sigma + \boldsymbol\sigma (\partial_\omega \mathbf{A}^\intercal). 
\end{align}
\end{subequations}
In the derivation of Eq.~(\ref{eq:eq_pwr}), the Wiener increment in Eq.~(\ref{eq:rsigmaeqs1}) should be replaced by $dw = dy_t - \sqrt{2\eta\kappa}\mathbf{v}^\intercal\mathbf{r}dt$, and the properties $\partial_\omega(dy_t) = 0$ has been used. We then apply the It\^o rule for two stochastic processes $f$ and $g$, 
\begin{equation}\label{eq:Itorule}
d(fg) = gdf + fdg + dfdg, 
\end{equation}
and derive the following differential relations
\begin{equation}
\begin{aligned}
d\mathbb{E}[\mathbf{r}\mathbf{r}^\intercal] &= \mathbb{E}[(d\mathbf{r})\mathbf{r}^\intercal] + \mathbb{E}[\mathbf{r}(d\mathbf{r}^\intercal)] + \mathbb{E}[(d\mathbf{r})(d\mathbf{r}^\intercal)], \\
d\mathbb{E}[(\partial_\omega\mathbf{r})(\partial_\omega\mathbf{r}^\intercal)] &= \mathbb{E}[(d\partial_\omega\mathbf{r})(\partial_\omega\mathbf{r}^\intercal)] + \mathbb{E}[(\partial_\omega\mathbf{r})(d\partial_\omega\mathbf{r}^\intercal)] \nonumber\\
&\displaystyle~~~+ \mathbb{E}[(d\partial_\omega\mathbf{r})(d\partial_\omega\mathbf{r}^\intercal)], \\
d\mathbb{E}[(\partial_\omega\mathbf{r})\mathbf{r}^\intercal] &= \mathbb{E}[(d\partial_\omega\mathbf{r})\mathbf{r}^\intercal] + \mathbb{E}[(\partial_\omega\mathbf{r})d\mathbf{r}^\intercal] \nonumber\\
&~~~ + \mathbb{E}[(d\partial_\omega\mathbf{r})d\mathbf{r}^\intercal]. 
\end{aligned}
\end{equation}
Finally, substituting Eqs.~(\ref{eq:rsigmaeqs}) and (\ref{eq:eqs_pwrSigma}) into these relations leads to the promised set of deterministic differential equations
\begin{equation}
\renewcommand{\arraystretch}{2}
\begin{aligned}
&\displaystyle \frac{d\mathbb{E}[\mathbf{r}\mathbf{r}^\intercal]}{dt} = \mathbf{A} \mathbb{E}[\mathbf{r}\mathbf{r}^\intercal] + \mathbb{E}[\mathbf{r}\mathbf{r}^\intercal]\mathbf{A}^\intercal + \frac{\kappa\eta}{2}(\boldsymbol\sigma-\mathbf{I})\mathbf{B}(\boldsymbol\sigma-\mathbf{I}), \\
&\displaystyle \frac{d\mathbb{E}[(\partial_\omega \mathbf{r}) (\partial_\omega \mathbf{r}^\intercal)]}{dt}  = (\partial_\omega\mathbf{A}) \mathbb{E}[\mathbf{r}(\partial_\omega \mathbf{r}^\intercal)] + \mathbf{A}\mathbb{E}[(\partial_\omega\mathbf{r})(\partial_\omega \mathbf{r}^\intercal)] \nonumber\\
&\displaystyle~~~~~~~~~~ - \kappa\eta(\boldsymbol\sigma-\mathbf{I})\mathbf{B}\mathbb{E}[(\partial_\omega\mathbf{r})(\partial_\omega \mathbf{r}^\intercal)]  + \frac{\kappa\eta}{2}  (\partial_\omega\boldsymbol\sigma)\mathbf{B} (\partial_\omega\boldsymbol\sigma) \\
&\displaystyle~~~~~~~~~~ + \mathbb{E}[(\partial_\omega \mathbf{r})\mathbf{r}^\intercal](\partial_\omega\mathbf{A}^\intercal)  + \mathbb{E}[(\partial_\omega \mathbf{r})(\partial_\omega\mathbf{r}^\intercal)]\mathbf{A}^\intercal  \nonumber\\
&\displaystyle~~~~~~~~~~ - \kappa\eta\mathbb{E}[(\partial_\omega \mathbf{r})(\partial_\omega\mathbf{r}^\intercal)]\mathbf{B}(\boldsymbol\sigma-\mathbf{I}), \\
&\displaystyle \frac{d\mathbb{E}[(\partial_\omega \mathbf{r}) \mathbf{r}^\intercal]}{dt}  = (\partial_\omega\mathbf{A}) \mathbb{E}[\mathbf{r}\mathbf{r}^\intercal] + \mathbf{A}\mathbb{E}[(\partial_\omega\mathbf{r})\mathbf{r}^\intercal] \nonumber\\
&\displaystyle~~~~~~~~~~ - \kappa\eta(\boldsymbol\sigma-\mathbf{I})\mathbf{B}\mathbb{E}[(\partial_\omega\mathbf{r})\mathbf{r}^\intercal] + \mathbb{E}[(\partial_\omega \mathbf{r})\mathbf{r}^\intercal] \mathbf{A}^\intercal \nonumber\\
&\displaystyle~~~~~~~~~~ + \frac{\kappa\eta}{2}(\partial_\omega\boldsymbol\sigma)\mathbf{B}(\boldsymbol\sigma-\mathbf{I}). \\
\\
\\
\\
\\
\end{aligned}
\end{equation}
These differential equations can then be integrated to obtain the solution for the integrand in Eq.~(\ref{eq:integrand}).

\bibliography{refs}

\end{document}